\documentclass{article}
\usepackage{spconf,amsmath,graphicx}
\usepackage{float}
\usepackage{wrapfig}
\usepackage{subcaption}

\usepackage{amsmath}
\usepackage{graphicx}
\usepackage[colorinlistoftodos]{todonotes}
\usepackage[colorlinks=true, allcolors=blue]{hyperref}
\usepackage{newfloat}
\DeclareFloatingEnvironment[placement={!ht},name=List]{mylist}
\usepackage{gensymb}
\usepackage{siunitx}

\usepackage{mathtools}

\DeclarePairedDelimiter\ceil{\lceil}{\rceil}
\DeclarePairedDelimiter\floor{\lfloor}{\rfloor}

\title{On the Brain Networks of Complex Problem Solving}

\name{ Abdullah Alchihabi $^{\dagger}$ \enskip Omer Ekmekci $^{\dagger}$ \enskip Baran B. Kivilcim $^{\dagger}$ \enskip Sharlene D. Newman$^{\star}$ \enskip Fatos T. Yarman Vural  $^{\dagger}$}

\address{ $^{\dagger}$ 
Department of Computer Engineering\\
Middle East Technical University,  \{abdullah.alchihabi, oekmekci, baran.kivilcim, vural\}@ceng.metu.edu.tr \\
       $^{\star}$ 
      Department of Psychological and Brain Sciences \\
Indiana University , sdnewman@indiana.edu }

\begin{document}

\maketitle
\begin{abstract}

Complex problem solving is a high level cognitive process which has been thoroughly studied over the last decade. The Tower of London (TOL) is a task that has been widely used to study problem-solving. In this study, we aim to explore the underlying cognitive network dynamics among anatomical regions of complex problem solving and its sub-phases, namely \textit{planning} and \textit{execution}. A new brain network construction model establishing dynamic functional brain networks using fMRI is proposed. The first step of the model is a preprocessing pipeline that manages to decrease the spatial redundancy while increasing the temporal resolution of the fMRI recordings. Then, dynamic brain networks are estimated using artificial neural networks. The network properties of the estimated brain networks are studied in order to identify regions of interest, such as hubs and subgroups of densely connected brain regions. The major similarities and dissimilarities of the network structure of planning and execution phases are highlighted. Our findings show the hubs and clusters of densely interconnected regions during both subtasks. It is observed that there are more hubs during the planning phase compared to the execution phase, and the clusters are more strongly connected during planning compared to execution.

\end{abstract}

\begin{keywords}
fMRI, Machine Learning, Brain Networks, Tower of London, Complex Problem Solving.
\end{keywords}

\section{Introduction}

Complex problem solving has been the focus of numerous studies in the field of neuroscience for over 30 years given the large number of high-level cognitive tasks that fall under its umbrella. To name a few, complex problem solving includes, strategy formation, coordination, sequencing of mental functions, and holding information on-line. These complex high-level cognitive sub-processes make revealing the inner workings of problem solving difficult.

The Tower of London (TOL) task, designed by Tim Shallice in (1982) \cite{shallice1982specific}, has been one of the standard tools in the literature to study complex problem solving. It consists of three bins having different capacities with colored balls placed in the bins; the aim is to rearrange the balls from their initial state to a predetermined goal state while moving one ball at a time and taking into consideration the limited capacity of each bin. 

TOL also has been used to investigate the effect of various clinical disorders on functions associated with the prefrontal cortex such as planning. It has been utilized to identify executive dysfunction in children and adolescents suffering from epilepsy and seizures \cite{MacAllister}. It also has been used to analyze the cognitive activation patterns of individuals suffering from depression \cite{Goethals}. Additionally, it has been employed to examine cognitive impairment in patient's diagnosed with Parkinson disease \cite{Rektorova}. In another study, TOL along with fMRI have been employed to study the differences in the neural basis of planning and executive function between first-episode schizophrenia patients and healthy subjects \cite{Rasser}.

Besides clinical disorders, TOL has also been employed to study the effect of various parameters on complex problem solving performance in healthy subjects. The predictive power of working memory, inhibition, and fluid intelligence on TOL performance has been explored \cite{Unterrainer2004,Zook2004}. Also, the effect of physical activity and exercise, age, gender, and impairment in the executive function on planning and problem-solving ability and its underlying neural basis have all been studied \cite{Chang,Albert,Zook2006,Boghi,unterrainer2005influence,kaller2012linking,desco2011mathematically}. 
 
TOL itself has many variations due to its large number of parameters such as goal hierarchy, demand for subgoal generation, number of solution paths, and the existence of suboptimal alternatives. Several studies have examined the effect of the aforementioned structural parameters along with numerous other non-structural parameters including instructions, experience, environment and problem-solving strategy on the performance of subjects when solving TOL puzzles, where performance is measured by pre-planning time and accuracy \cite{Kaller2004,Berg,Unterrainer2005,newman2007tower,Unterrainer2003,Campbell,Newman2010}.

Despite the popularity of the TOL problem in the literature and the wide range of problems covered in these studies, relatively few researchers have explored the underlying network structures. In \cite{Newman2009}, the involvement of the parietal cortex, prefrontal cortex, basal ganglia and anterior cingulate in complex problem solving was reported. The activation patterns of the dorsolateral and rostrolateral subregions of the prefrontal cortex during planning has been examined \cite{Wagner}. The focus of another study has been the hemispheric differences in the pre-frontal cortex during planning and execution as well as the contribution of the superior parietal region to spatial working memory \cite{Newman2003}. In addition, some work has been done to investigate the variance in the neural basis of planning between standard and expert subjects \cite{Cazalis}. Other works have investigated the cognitive activation patterns during planning and execution subtasks \cite{van2003frontostriatal,spreng2010default,kipping2018trade,beauchamp2003dynamic}. Given this brief literature review, a holistic understanding of the active anatomical brain regions, their respective roles and their interactions during complex problem-solving is an important, yet lacking research study. 

Numerous studies have proposed various computational models in order to build brain networks from fMRI measurements, both during cognitive tasks or during resting state. These studies represent a shift in the literature towards brain decoding algorithms that are based on the connectivity patterns in the brain motivated by the findings that these patterns provide more information about cognitive tasks than the isolated behavior of individual voxel groups or anatomical regions \cite{shirer2012decoding,ekman2012predicting,onal2017new,lindquist2008statistical,richiardi2013machine}. 

Some of these studies focused on the pairwise relationships between voxels or brain regions. For example, Pearson correlation has been used in order to construct undirected functional connectivity graphs at different frequency resolutions in \cite{richiardi2011decoding}. Also, pairwise correlations and mutual information have also been used in order build functional brain networks in various studies aiming to investigate the network differences between patients with Schizophrenia or Alzheimer’s disease and healthy subjects \cite{lynall2010functional,menon2011large,kurmukov2017classifying}. 
Partial correlation along with constrained linear regression was also used to generate brain networks in \cite{lee2011sparse}. 
 
Other studies take advantage of the locality property of the brain by constructing local mesh networks around each brain region then representing the entire brain network as an ensemble of local meshes. In such studies, the Blood-Oxygenation Level Dependent (BOLD) response of each brain region is estimated as a linear combination of the responses of its closest neighboring regions. Levinson-Durbin recursion has been applied in several studies in order to estimate the edge weights of each local star mesh, where the nodes are the neighboring regions of the seed brain region \cite{firat2013functional,alchihabi2018dynamic}.
Ridge regression has also been used to estimate edge weights while constructing local mesh networks across windows of time \cite{onal2015modeling,onal2017new}.

Other works suggested various methods to prune the constructed brain networks. In \cite{wee2012constrained}, researchers established resting-state brain networks as sparse constrained networks using both L1 and L2 regularization to introduce sparsity and control for across-subject variability. Another study used a two-step model to build functional brain networks, where at first a sparse multivariate autoregressive model was employed with penalized regression to estimate the brain networks. Then, false discovery rate (FDR) is used to prune low probability connections to present sparsity in the brain network \cite{valdes2005estimating}. 


After constructing brain networks, the statistical properties of the established networks are studied in order to obtain neuroscientific insights related to the experiment at hand.
Several measures of centrality have been proposed that aim to identify potential hubs that are central to the flow of information in the network such as: node degree, node strength and node betweenness centrality. In addition, measures of functional segregation have been proposed that aim to detect subgroups or modules of densely interconnected anatomic regions such as: local efficiency, clustering coefficient and transitivity \cite{rubinov2010complex,park2013structural,power2010development}. Furthermore, the node degree distribution of constructed brain networks has been measured and compared with power law and truncated power law distributions \cite{bassett2006small}. Also, the small-world property of brain networks has been studied extensively in numerous studies \cite{bassett2006small,park2013structural}. Some studies, such as \cite{rubinov2011weight}, have further extended the literature by defining the null model for weighted undirected functional brain networks. Further work have focused on controlling for family-wise error (FWE) that complements false discovery rate (FDR) \cite{zalesky2010network}.

Several studies have compared the network properties of functional brain networks across different age groups \cite{achard2007efficiency,power2010development} and under different developmental factors \cite{blesa2018early}. Other studies have performed similar analysis to compare the network properties of healthy individuals against those suffering from several diseases related to cognitive impairment (Alzheimer, epilepsy, Schizophrenia) \cite{braun2015human}.

In this study, we propose a dynamic functional brain network model, extracted from fMRI measurements using artificial neural networks. The decoding power of the suggested brain network model is examined by distinguishing the two phases of complex problem solving, namely, planning and execution.
Then, the network properties of the dynamic brain networks are studied in order to identify the active anatomical regions during both planning and execution phases of complex problem solving. Potential hubs and clusters of densely connected brain regions are identified for both subtasks. Furthermore, the distinctions and similarities between 	planning and execution networks are highlighted. The results identify both active, inactive, hub regions as well as clusters of densely connected anatomical regions during complex problem task. In addition, results show that there are potential hubs during planning phase compared to execution phase, also the clusters of densely interconnected regions are significantly more strongly connected during planning compared to execution.

\section{TOL Experiment Procedure}

In this section, we introduce the details of the experiment as well as data collection and preprocessing methods.

\subsection{Participants \& Stimuli}

18 college students aged between 19 and 38 participated in the experiment after signing informed, written consent documents approved by the Indiana University Institutional Review Board. The subjects solved a computerized version of TOL problem, two configurations were presented at the beginning of each puzzle: the initial state and the goal state.
The subjects were asked to transform the initial state into the goal state using the minimum number of moves. However, the subjects were not informed of the minimum number of moves needed to solve a given puzzle nor of the existence of multiple solution paths.

\subsection{Procedure}

Each subject underwent a practice session before entering the scanning session to acquaint subjects with the TOL problem. The subjects were given the following instructions: “You will be asked to solve a series of puzzles. The goal of the puzzle is to make the ‘start’ or ‘current’ state match the ‘goal’ state (They were shown an example). Try to solve the problems in the minimum number of moves by planning ahead. Work as quickly and accurately as possible, but accuracy is more important than speed.” 

The scanning session consisted of 4 runs, each run included 18 timed puzzles, with a 5 second planning only time slot during which subjects were not allowed to move the balls. However, they were allowed to continue planning after the 5 seconds planning only time slot if they chose to do so. Following every puzzle, there was a 12-second rest period where subjects focused on a plus sign in the center of the screen. Each run was also followed by a 28-second fixation period.

\subsection{fMRI Data Acquisition \& Preliminary Analysis}

The fMRI images were collected using a 3 T Siemens TRIO scanner with an 8-channel radio frequency coil located in the Imaging Research Facility at Indiana University. The images were acquired in 18 5 mm thick oblique axial slices using the following set of parameters: TR=1000 ms, TE=25 ms, flip angle=\ang{60}, voxel size=3.125 mm$\times$3.125 mm$\times$5 mm with a 1 mm gap.

The statistical parametric mapping toolbox was used to perform the preliminary data analysis that included: image correction for slice acquisition timing, resampling, spatial smoothing, motion correction and normalization to the Montreal Neurological Institute (MNI) EPI template. Further details concerning the procedure and data acquisition can be found in \cite{Newman2009}.

\section{Mesh Brain Networks for Complex Problem Solving}

\subsection{Data Processing}

Given the small number of subjects in the TOL dataset, and the large number of voxels in each brain volume (185,405 voxels per time instant), voxel selection is used to reduce the spatial resolution of the collected brain images and dampen the noise that is inherent in the data. Furthermore, due to the short duration of each puzzle (max 15 seconds) and the relatively low sampling rate (TR = 1 sec), temporal interpolation is needed in order to increase the number of brain volumes for each puzzle. Finally, Gaussian noise is used in order to regularize the data and improve generalization.

\subsubsection{Voxel Selection}

First, an ANOVA feature selection method is used to choose the most discriminative subset of voxels and discard the remaining ones \cite{Afrasiyabi1,Pereira,Cox}. For this purpose, we calculate the $f$-value score of each voxel $\boldsymbol{v_i}$ as shown in equation \ref{eq:F-Value}:

\begin{equation}
f\_score_i = \frac{MSB(\boldsymbol{v_i}, \boldsymbol{y_{label}})}{MSW(\boldsymbol{v_i}, \boldsymbol{y_{label}})}
\label{eq:F-Value}
\end{equation}

where $\boldsymbol{y_{label}}$ is the label indicating the subtask (Planning or Execution). $MSB(\boldsymbol{v_i}, \boldsymbol{y_{label}})$ is the mean square value between voxel $i$ and the label vector $\boldsymbol{y_{label}}$ which is calculated by equation \ref{eq:MSB}

\begin{equation}
MSB(\boldsymbol{v_i}, \boldsymbol{y_{label}}) = \frac{SSB(\boldsymbol{v_i}, \boldsymbol{y_{label}})}{df_{between}}
\label{eq:MSB},
\end{equation}

$SSB(\boldsymbol{v_i}, \boldsymbol{y_{label}})$ is the sum of squares between $\boldsymbol{y_{label}}$ and $\boldsymbol{v_i}$, $df_{between}$ is the number of groups minus one.\\
$MSW(\boldsymbol{v_i},\boldsymbol{y_{label}})$ is the mean square value within voxel $i$ and the label vector $\boldsymbol{y_{label}}$ and it is calculated by \ref{eq:MSW} 

\begin{equation}
MSW(\boldsymbol{v_i},\boldsymbol{y_{label}}) = \frac{SSW(\boldsymbol{v_i},\boldsymbol{y_{label}})}{df_{within}}
\label{eq:MSW}
\end{equation}

where, $SSW(\boldsymbol{v_i},\boldsymbol{y_{label}})$ is the sum of squares within group and $df_{within}$ is the degree of freedom within (total number of elements in $\boldsymbol{v_i}$ and $\boldsymbol{y_{label}}$ minus the number of groups).

We order the voxels according to their $f$-value scores. Then, the distribution of $f$-value scores of all voxels is plotted in order to determine the appropriate number of voxels to retain. Voxel selection is applied to the voxels of all brain regions except the ones located in the cerebellum, which we exclude during network extraction.

Voxel selection successfully manages to significantly reduce the number of voxels in each brain volume, thus, making the space and time complexity of the analysis on the dataset feasible given the large total number of voxels in each brain volume, 185,405 voxels per time instant. It is also a necessary step for decreasing the curse of dimensionality problem for decoding the planning and execution phases.  

The BOLD response of the selected voxels is then averaged into their corresponding brain regions defined by the automated anatomical labeling (AAL) atlas \cite{AAL} as shown in equation \ref{eq:AAL-Averaging}:

\begin{equation}
\boldsymbol{r_j}  = \frac{\sum_{i \in \zeta[j]} \boldsymbol{v_i} }{| \zeta[j] |}
\label{eq:AAL-Averaging}
\end{equation}

where $\boldsymbol{r_j}$ is the BOLD response of region $j$, $\boldsymbol{v_i}$ is the BOLD response of voxel $i$ and $\zeta[j]$ is the set of selected voxels located in region $j$.
Averaging the selected voxels in their corresponding anatomical regions smooths the noise embedded in fMRI signal to a certain degree, and further reduces the dimensionality of each brain volume. As a result, each region is represented by a BOLD response, thus, enabling us to investigate the role and contribution of each region to the planning and execution phases of the problem solving task. 

\subsubsection{Interpolation} 

It is well-known that in spite of its high spatial resolution, fMRI signal has very low temporal resolution compared to EEG signal. In this study, we interpolate the fMRI signal in order to compensate for this drawback and study the effect of interpolation on decoding planning and execution phases of TOL. 

In the TOL study subjects solved a puzzle in at most 15 seconds and the sampling rate, TR, is 1000 ms. Interpolation is used to increase the temporal resolution by estimating $z$ extra brain volumes between each two consecutive measured brain volumes. As a result, the total number of available brain volumes for each puzzle becomes $ n + z * (n-1)$, where $n$ is the number of measured brain volumes of a given puzzle. We use the cubic spline interpolation function rather than linear interpolation methods in order to prevent edge effects and smoothing out the spikes between the measured brain volumes \cite{mckinley1998cubic}. 



In order to analyze the effect of time interpolation and to estimate an acceptable number of inserted brain volumes $z$, we compare the Fourier Transform of the fMRI signal computed before and after interpolation so that the frequency content of the signal is not distorted by interpolation. The original single-sided amplitude of the signal and the one obtained after interpolation are compared in order to ensure that interpolation is preserving the smooth peaks of the data in the frequency domain \cite{frigo1998fftw,cochran1967fast}.

\subsubsection{Injecting Gaussian Noise }

When modeling a deterministic signal by a probabilistic method, adding noise to the signal decreases the estimation error in most of the practical applications. The final phase of preprocessing is adding a Gaussian noise to the interpolated time series of the BOLD response in each anatomical region. For this purpose, instead of just injecting white noise, a rather informed noise, \textit{colorful Gaussian noise}, is added. In order to reflect the corresponding brain region's properties, for each sample the additive noise sample is generated from a Gaussian distribution having mean and variance of that anatomical region. This newly generated samples not only act like a natural regularizer to improve generalization performance of brain decoding, but, also help making local mesh estimation algorithms more stable when generating brain networks \cite{matsuoka1992noise,reed1992regularization}. 

Given a representative time series from a particular brain region, $i$ represents the index of an anatomical region. The new samples are generated with vector addition of noise while preserving the signal-to-noise ratio (SNR) as in $\boldsymbol{\tilde{r}_{j}}  = \boldsymbol{r_{j}}  + \boldsymbol{\tau_{j}} $, where $\boldsymbol{\tau_{j}} $ is a noise vector sampled from $ \mathcal{ N } ( \alpha_{noise} \;  \mu( \boldsymbol{r_{j}}),\, \beta_{noise}  \; \sigma^{2}( \boldsymbol{r_{j}})) $, $\alpha_{noise}$ and $\beta_{noise}$ are the scaling factors which are set empirically, to optimize the decoding performance. 

\subsection{Building Dynamic Brain Network with Neural Networks}

After applying the preprocessing pipeline, we construct dynamic functional brain networks. 
In order to do that, we partition each time series, which represents an anatomical region into fixed-size windows, where each window, $win(t)$, is centered at the measured brain volume at time instance, $t$. The size of each window is $ Win\_Size =  z  + 1 $ brain volumes, where $z$ is the number of interpolated brain volumes in each window. Equation \ref{eq:Window_size} shows the time instances included in each window.

\begin{equation}
\boldsymbol{win(t)} =  \Bigg[ t - \floor*{\frac{z}{2}} ,..  ,  t, .. ,  t + \ceil*{\frac{z}{2}} \Bigg]
\label{eq:Window_size}
\end{equation}

We construct a dynamic brain network, $N(t) = (V,W(t))$, for each time window $win(t)$,
where $V$ is the set of nodes of the graph corresponding to the brain anatomical regions, while $W(t) = \{w_{t,j,i}  |  \forall i , j \in V \}$ is the directed weighted edges between the nodes of the graph within time window $win(t)$. The nodes of the graph are the AAL defined brain regions\cite{AAL}, except for the regions located in the cerebellum. The nodes are then pruned using voxel selection, as some anatomical regions contribute no voxels at all and get deleted from the set of nodes of the graph $V$. 

Note that our aim is to label the BOLD responses measured at each brain volume as it belongs to one of the two phases of complex problem solving, namely, planning and execution. For this purpose, we represent each brain volume measured at a time instant $t$ by a network, which shows the relationship among the anatomical regions. This dynamic network representation will allow us to investigate the network properties of planning and execution subtasks. In this section, we describe how we estimate the weights of the edges, $W(t)$, of the brain network, $N(t)$, for each time instance, $t$, where we employ the method proposed in \cite{kivilcim2018modeling}.

For each window $\boldsymbol{win(t)}$, we define the functional neighborhood matrix, $\Omega_{t}$. The entries of $\Omega_{t}$ are binary, either 1 or 0, indicating if there is a connection between two regions or not. The size of the matrix is $MxM$, where $M$ is  the number of brain anatomical regions. The functional neighborhood matrix contains no self-connections, thus, $\Omega_{t}(i,i) = 0 \, \forall i \in [1,M] $. Also, the brain regions pruned by voxel selection contributing no voxels have no in/out connections, thus, the corresponding entries in $\Omega_{t}$ are all zeros. The connectivity of each region to the rest of the regions is determined using Pearson correlation, as follows: first, for every region $i$, we measure the Pearson correlation between its BOLD response $\boldsymbol{r_{i,t}}$ and the BOLD responses of all the other remaining regions as shown below: 

\begin{center}
\begin{equation}\label{eq:Pearson}
cor(\boldsymbol{r_{i,t}},\boldsymbol{r_{j,t}}) = \frac{cov(\boldsymbol{r_{i,t}},\boldsymbol{r_{j,t}})}{\sigma(\boldsymbol{r_{i,t}}) \sigma(\boldsymbol{r_{j,t}})} , 
\end{equation}
\end{center}

where $\boldsymbol{r_{i,t}}$ is the BOLD response of region $i$ across time window $win(t)$, $cov(\boldsymbol{r_{i,t}},\boldsymbol{r_{j,t}})$ is the covariance between the corresponding BOLD responses of regions $i$ and $j$. $\sigma$ is the standard deviation of the BOLD response of a given region. Thus, the higher the Pearson correlation between two regions the closer they are to each other in the functional neighborhood system.

Then, we select $p$ of the regions with the highest correlation scores with region $i$. Thus, obtaining the neighborhood set $\eta_{p}[i]$, which contains the $p$ closest brain regions to region $i$. Finally, we define the $\Omega_{t}(i,j)$ as the connectivity between the regions $i$ and $j$, using the constructed neighborhood sets as follows: 

\begin{equation}
\Omega_{t}(i,j) = 
\begin{cases}
1, & \text{if}\ j \in \eta_{p}[i]    \\
0,  & \text{otherwise} . \\ 
\end{cases}
\label{eq:Neighborhood_matrix_pearson}
\end{equation}

Note that each anatomical region is connected to its $p$ closest functional neighbors. This approach forms a star mesh around each anatomical region. The ensemble of all of the local meshes creates a brain network at each time instance. Note, also, that Pearson correlation values are not used as the weights between two regions. They are just used to identify the nodes of each local mesh formed around an anatomical region. The estimated brain network becomes sparser as $p$ gets smaller. When $p$ is set to the number of anatomic regions, $M$, the network becomes fully connected. The selection method for the degree of neighborhood, $p$, is explained in the next section. This approach of defining the connectivity matrix not only makes the network representation sparse for small $p$ values, but, it also constructs a network which is connected in functionally closest regions, satisfying the locality property of the human brain. 

After having determined the edges of the brain graph using the functional neighborhood matrix $\Omega_{t}$, all that is left is to estimate the weights of these edges at each local mesh. In order to do that, we represent the response of each region $i$ ($\boldsymbol{r_{i,t}}$) as a linear combination of its closest $p$-functional neighbors as shown in equation \ref{eq:Mesh_Linear_Combination}, 

\begin{center}
\begin{equation}\label{eq:Mesh_Linear_Combination}
\boldsymbol{\hat{r}_{i,t}} =\sum_{j \in \eta_{p}[i]} w_{t,j,i}  \boldsymbol{r_{j,t}}  + \epsilon_{i,t} .
\end{equation}
\end{center}

In equation \ref{eq:Mesh_Linear_Combination}, $\boldsymbol{\hat{r}_{i,t}}$ represents the representative time series of the BOLD response of region $i$ within the time window $win(t)$, $w_{t,j,i}$ is the estimated  edge weight between node (region) $i$ and $j$ at time instance $t$. $\eta_{p}[i]$ is the $p$ closest functional neighbors of region $i$. \\

Onal et.al. \cite{Onal16} estimated the arc-weights for each mesh formed around region $i$ for each time window $win(t)$ by minimizing the mean-squared error loss function using Ridge regression. In this approach, the mean-squared error loss function is minimized with respect to $w_{t,j,i}$, for each region, independent of the other regions, where the expectation is taken over the time-instances, in window $win(t)$ as shown in equation \ref{eq:Loss_General} . 

\begin{center}
\begin{equation}\label{eq:Loss_General}
E[(\epsilon_{i,t})^2]=E[(\boldsymbol{\hat{r}_{i,t}}  - \sum_{j \in \eta_{p}[i]} w_{t,j,i} \boldsymbol{r_{j,t}}   )^2  ] +  \lambda ||  w_{t,j,i} ||^2 ,
\end{equation}
\end{center}

where $\lambda$ is the L2 regularization parameter whose value is optimized using cross-validation. L2 regularization is used in order to improve the generalization of the constructed mesh networks. Note that the estimated arc-weights, $w_{t,j,i} \neq  w_{t,i,j}$. Therefore, the ensemble of meshes yields a directed brain network.

In this work, we define an artificial neural network to estimate the values of mesh arc-weights for all anatomical regions jointly in each time window, as proposed in \cite{kivilcim2018modeling}. In this method, we estimate the mesh arc-weights matrix $W(t) = \{w_{t,j,i} | j , i \in V\}$ using a feedforward neural network. The architecture of this network consists of an input layer and an output layer, both containing $M$ nodes corresponding to each brain region. The edges of this network are constructed using the neighborhood matrix $\Omega_{t}$. There is an edge between node $i$ of the output layer and node $j$ from the input layer, if $\Omega_{t}(i,j) = 1$.

The loss function of the suggested artificial neural network is given in equation \ref{eq:Loss_NN}, where $W$ is the weight matrix of the entire neural network that corresponds to directed edge weights of the brain graph and $W_{i}$ is the row of matrix $W$ corresponding to region $i$: 

\begin{center}
\begin{equation}\label{eq:Loss_NN}
\begin{split}
Loss(Output_i) & =  E[(\epsilon_{i,t})^2]   +  \lambda W_{i}^T W_{i} \\ 
			   & = E[( \boldsymbol{r_{i,t}}   - \sum_{j \in \eta_{p}[i]} w_{t,j,i} \boldsymbol{r_{j,t}} )^2 ] +  \lambda W_{i}^T W_{i} .
\end{split}
\end{equation}
\end{center}

We train the aforementioned artificial neural network in order to obtain the weights of the brain network at each time instance $t$ that minimize the loss function by applying a gradient descent optimization method as shown in equation \ref{eq:Update_weights}, 

\begin{center}
\begin{equation}\label{eq:Update_weights}
w_{t,j,i}^{(\kappa)}  = w_{t,j,i}^{(\kappa-1)} - \alpha_{learning} \; \frac{\partial E[(\epsilon_{i,t})^2] }{\partial w_{t,j,i}}  ,
\end{equation}
\end{center}

where $w_{t,j,i}^{(\kappa)} $ is the weight of the edge from node $j$ to node $i$ at epoch (iteration) $\kappa$, $\alpha_{learning}$ is the learning rate. The number of epochs and learning rate used to train the network are optimized empirically using cross-validation.

Finally, the weights of the above artificial neural network, computed for each $win(t)$, correspond to the edge weights of the dynamic brain network, $N(t) = (V,W(t))$, at each time instant $t$. Thus, we refer to the brain networks using their window indices in order to obtain a set of dynamic brain networks $T=\{N(1),N(2), ... N({tot\_win})\}$, where $N(t)$ is the brain network for time window $win(t)$ and $tot\_win$ is the total number of time windows.

\subsection{Network Metrics for Analyzing Brain Networks}

In this section, we introduce some measures which we will use to investigate the network properties of each phase of the complex problem solving task, namely, planning and execution, using the estimated dynamic brain functional networks. The connectivity patterns of anatomical regions are analyzed by the set of network measures, given below. Two separate sets of measures are used, namely, measures of centrality and segregation. Since our estimated brain networks are directed, we distinguish the incoming and outgoing edges in the network, while defining the measures.

Recall that the suggested brain network $N(t) = (V,W(t))$ consist of a set of nodes, $V$, each of which corresponding to one of the $M$ anatomical regions. $W(t)$ is the dynamic edge weight matrix with the entries, $w_{i,j}$, representing the weight of the edge from node $i$ to node $j$. For the sake of simplicity, we omit the time dependency parameter $t$, since we compute the network properties at each time instant. Matrix $A$ is the binarized version of $W(t)$ matrix, where $a_{i,j}$ takes value 0 if $(w_{i,j} == 0 )$ and takes value 1 otherwise. 
 
\subsubsection{Measures of Centrality}

Measures of centrality aim to identify brain regions that play a central role in the flow of information in the brain network, or nodes that can be identified as hubs. It is commonly measured using node degree, node strength and node betweenness centrality, which are defined below.

\subsubsection{Node Degree }
The degree of a node is the total number of its edges as shown in equation \ref{eq:node_degree}, where $degree_i$ is the degree of node $i$, $V$ is the set of all nodes in the graph and $a_{i,j}$ is the edge between node $i$ and node $j$.

\begin{center}
\begin{equation}\label{eq:node_degree}
degree_i = \sum_{j \in V} a_{i,j}
\end{equation}
\end{center}

In the case of a directed graph, we distinguish two different metrics: node in-degree $degree_{i}^{in}$ and node out-degree $degree_{i}^{out}$ metrics which are shown in equations \ref{eq:node_indegree} and \ref{eq:node_outdegree} respectively where $a_{j,i} = 1$, if there is a directed edge from node $j$ to node $i$. 

\begin{center}
\begin{equation}\label{eq:node_indegree}
degree_{i}^{in} = \sum_{j \in V} a_{j,i}
\end{equation}
\end{center}

\begin{center}
\begin{equation}\label{eq:node_outdegree}
degree_{i}^{out} = \sum_{j \in V} a_{i,j}
\end{equation}
\end{center}

Node degree is a measure of centrality of the given nodes, where it aims to quantify the hub brain regions interacting with a large number of brain regions. Thus, a node with high degree indicates its central role in the network.

\subsubsection{Node Strength}
 
Node strength is the sum of the weights of edges connected to a given node \ref{eq:node_strength}, where $w_{i,j}$ is the weight of the edge between node $i$ and node $j$.

\begin{center}
\begin{equation}\label{eq:node_strength}
strength_{i} = \sum_{j \in V} w_{i,j}
\end{equation}
\end{center}

Similar to node degree, node strength, also, distinguishes two metrics in the case of directed graphs, namely, node in-strength $strength_{i}^{in}$ and out-strength $strength_{i}^{out}$ shown in equations \ref{eq:node_in_strength} and \ref{eq:node_out_strength} respectively, where $w_{j,i}$ is the weight of the edge from node $j$ to node $i$.

\begin{center}
\begin{equation}\label{eq:node_in_strength}
strength_{i}^{in} = \sum_{j \in V} w_{j,i} 
\end{equation}
\end{center}

\begin{center}
\begin{equation}\label{eq:node_out_strength}
strength_{i}^{out} = \sum_{j \in V} w_{i,j} .
\end{equation}
\end{center}

Node strength is a node centrality measure that is similar to node degree, which is used in the case of weighted graphs. Nodes with large strength values are tightly connected to other nodes in the network forming hub nodes.

\subsubsection{Node Betweenness Centrality}

Betweenness centrality of node $i$ is the fraction of the shortest paths in the network that pass through node $i$ as shown in equation \ref{eq:node_betweenness} 

\begin{center}
\begin{equation}\label{eq:node_betweenness}
betweenness_{i} =   \frac{1}{(M-1)(M-2)} \sum_{j,k \in  V} \frac{\rho_{j,k}^{i}}{\rho_{j,k}} \space ,
\end{equation}
\end{center}

where $\rho_{j,k}$ is the number of shortest paths betweens nodes $j$ and $k$, $\rho_{j,k}^{i}$ is the number of shortest paths between nodes $j$ and $k$ that pass through node $i$, nodes $i$, $j$ and $k$ are distinct nodes. 

Before measuring the betweenness centrality of a node, we need to change our perspective from connection weight matrix to connection length matrix since betweenness centrality is a distance-based metric. In connection weights matrix, larger weights imply higher correlation and shorter distance while it is the opposite in the case of length matrix. Connection length matrix is obtained by inverting the weights of the connection weight matrix. Then, the algorithm suggested in \cite{brandes2001faster} is employed in order to calculate the node betweenness centrality for each anatomical region.

Nodes with high betweenness centrality are expected to participate in many of the shortest paths of the networks. Thus, taking a crucial role in the information flow of the network. 

\subsubsection{Measures of Segregation}

Measures of segregation aim to quantify the existence of subgroups within brain networks, where the nodes are densely interconnected. These subgroups are commonly referred to as clusters or modules. The existence of such clusters in functional brain networks is a sign of interdependence among the nodes forming the cluster. Measures of segregation include clustering coefficient, transitivity and local efficiency. While global efficiency is a measure of functional integration representing how easy it is for information to flow in the network.

\subsubsection{Clustering Coefficient }

The clustering coefficient of a node $i$ is the fraction of triangles around node $i$ which is calculated by equation \ref{eq:clustering_coef} as proposed in \cite{fagiolo2007clustering}. It is defined as the fraction of the neighbors of node $i$ that are also neighbors of each other. 

\begin{center}
\begin{equation}\label{eq:clustering_coef}
C_i =   \frac{  \chi_{i} }{  [ (d_{i}^{out} + d_{i}^{in}) (d_{i}^{out} + d_{i}^{in} -1) - 2 \sum_{j  \in V }a_{i,j} a_{j,i} ]} .
\end{equation}
\end{center}

where $d_{i}^{in}$ is the in-degree of node $i$ and $d_{i}^{out}$ is the out-degree of node $i$. $\chi_i$ is the weighted geometric mean of triangles around node $i$ that is calculated by equation \ref{eq:nmb_triangles}. Recall that $a_{j,i} = 1$ , if there is a directed edge from node $j$ to node $i$ and $a_{j,i} = 0$, otherwise.

\begin{center}
\begin{equation}\label{eq:nmb_triangles}
\chi_i = \frac{1}{2}  \sum_{j,h \in V}  (w_{i,j} w_{i,h} w_{j,h})^{1/3} .
\end{equation}
\end{center}

The clustering coefficient of a node is the fraction of triangles around the node. It is defined as the fraction of the neighbors of the node that are also the neighbors of each other.

\subsubsection{Transitivity }

Transitivity of a node is similar to its clustering coefficient. However, transitivity is normalized over all nodes while cluster coefficient for each node is normalized independently which makes clustering coefficient biased towards nodes with low degree. Transitivity can be expressed as the ratio of triangles to triplets in the network.
It is calculated by equation \ref{eq:transitivity} , as suggested in \cite{fagiolo2007clustering}: 

\begin{center}
\begin{equation}\label{eq:transitivity}
T_i = \frac{  \chi_{i} }{  \sum_{j \in V}  [ (d_{j}^{out} + d_{j}^{in}) (d_{j}^{out} + d_{j}^{in} -1) - 2 \sum_{h  \in V } a_{j,h} a_{h,j}]} ,
\end{equation}
\end{center}
where $d_{j}^{in}$ is the in-degree of node $j$ and $d_{j}^{out}$ is the out-degree of node $j$. $\chi_i$ is the weighted geometric mean of triangles around node $i$ that is calculated by equation \ref{eq:nmb_triangles}. Note that $ a_{h,j} a_{j,h} = 1$ , if there exits  an  edge in both directions. 

\subsubsection{Global \& Local Efficiency }

The global efficiency of a brain network is a measure of its functional integration. It measures the degree of communication among the anatomical regions. Thus, it is closely related to the small-world property of a network.
Formally speaking, global efficiency is defined as the average of the inverse shortest path lengths between all pairs of nodes in the brain network. Equation \ref{eq:global_eff} shows how to calculate the global efficiency of a brain network, where $\varrho_{i,j}^{w}$ is the weighted shortest path length between two distinct nodes $i$ and $j$ \cite{rubinov2010complex}.

\begin{center}
\begin{equation}\label{eq:global_eff}
E_{global} = \frac{1}{M} \sum_{i \in V} \frac{\sum_{j \in V} (\varrho_{i,j}^{w})^{-1}}{M-1}
\end{equation}
\end{center}

On the other hand, the local efficiency of a network is defined as the global efficiency calculated over the neighborhood of a single node. The local efficiency is, thus, a measure of segregation rather than functional integration as it is closely related to clustering coefficient. While global efficiency is calculated for the entire network, local efficiency is calculated for each node in the network \cite{rubinov2010complex}.

\section{Experiments \& Results}

In this section, we explore the validity of the suggested network model by applying it to the TOL dataset. First, we analyze the effect of the preprocessing step on the brain decoding performance of planning and execution phases of complex problem solving.  Then, we investigate the validity of the dynamic functional brain network model proposed in this study. Finally, we analyze the network properties of the constructed functional brain networks for planning and execution subtasks.

\subsection{Voxel Selection} 

\begin{figure}[h]
\centering
\includegraphics[width=0.40\textwidth]{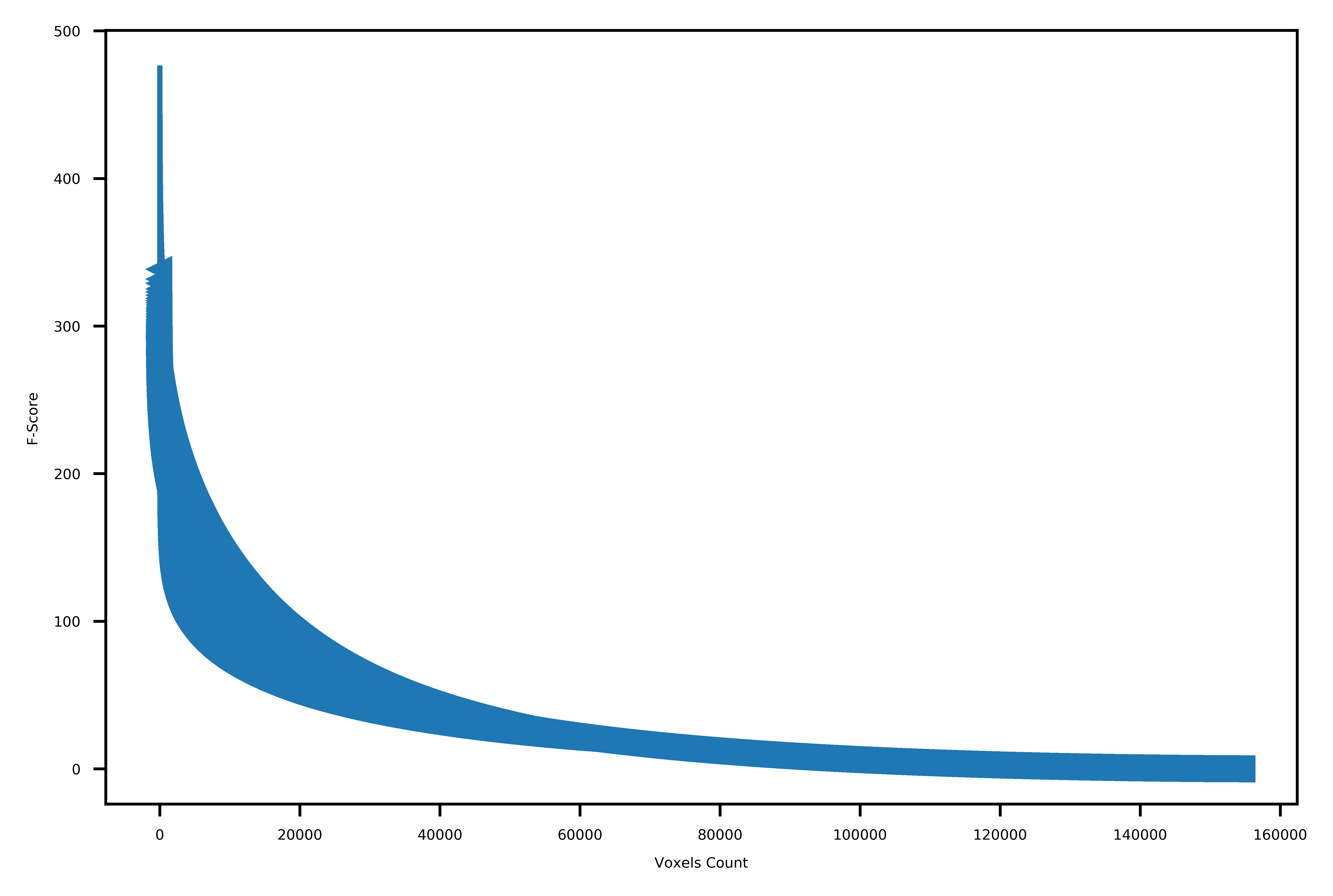}
\caption{Ordered $f$-scores of voxels.}
\label{fig:f_scores}
\end{figure}

At the first step of the proposed computational model, we discarded all of the voxels located in the cerebellum anatomical regions. Then, we calculated the $f$-score for each one of the remaining voxel and order the obtained $f$-scores of the voxels. Following that, we plotted the ordered $f$-scores of the voxels in order to determine the appropriate number of voxels to retain. Figure \ref{fig:f_scores} shows the ordered $f$-scores of the voxels averaged across all subjects. It can be observed from this figure that a relatively small number of voxels is crucial for discriminating the subtasks of problem solving while the remaining voxels do not have significant information concerning the subtasks of problem solving. Based on the $f$-score distribution shown in Figure \ref{fig:f_scores}, we kept the 10,000 voxels with the highest $f$-scores given the clear the elbow point whereas we discarded the remaining ones. 

\begin{figure*}[t]
		\centering
        \begin{subfigure}{\textwidth}
    	\centering
		\includegraphics[width=1\textwidth]{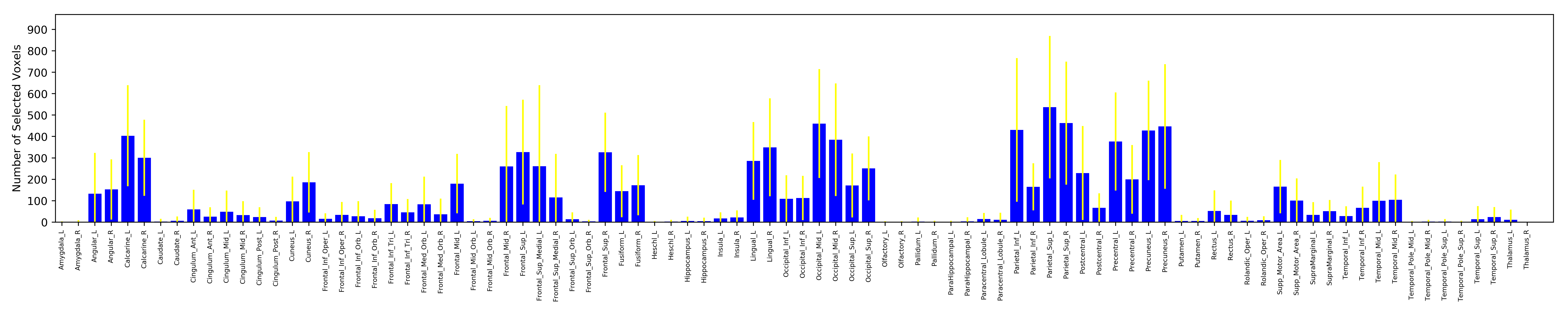} 
		\caption{Average number of voxels selected from each anatomical region across all subjects.}
  		\label{fig:Region_voxel_contribution}
        \end{subfigure}
        
        \centering
        \begin{subfigure}{\textwidth}
    	\centering
		\includegraphics[width=1\textwidth]{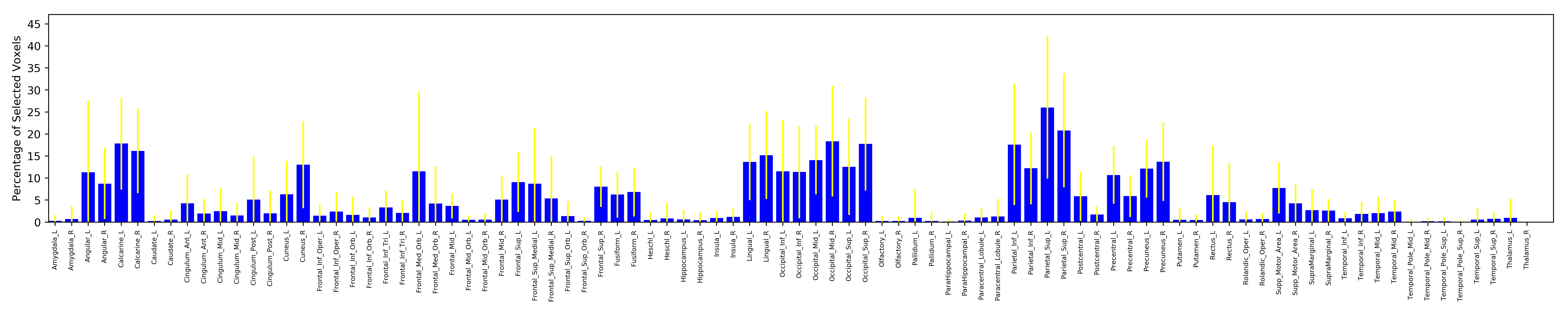} 
	\caption{Average percentage of voxels selected from each anatomical region across all subjects.}
  		\label{fig:Region_voxel_contribution_Precentage}
        \end{subfigure}
  				
        \caption{Distribution of selected voxels across anatomical regions, measured by number of selected voxels (top) and percentage of selected voxels (bottom) from each anatomical region.}

\end{figure*}

After selecting the 10,000 voxels with the highest $f$-scores of each session, we computed the number of selected voxels contained in each one of the 90 anatomical regions. We also calculated the percentage of selected voxels to the total number of voxels located in each anatomical region.   
Figure \ref{fig:Region_voxel_contribution} shows the average number of voxels contributed by each region across all subjects with its corresponding standard deviation, Figure \ref{fig:Region_voxel_contribution_Precentage} shows the average percentage of voxels contributed by each region across all subjects with its corresponding standard deviation.

It is clear from these figures that a large number of regions contribute little to no voxels, such as the amygdala, caudate, heschl gyrus, hippocampus, pallidum, putamen, temporal pole, superior temporal cortex, thalamus and parahippocampus. A small number of regions contribute a significantly large number of voxels (over 300 voxels each) during complex problem solving, such as occipital, precentral, precuneus and parietal regions. 

Furthermore, Figure \ref{fig:Region_voxel_contribution_Precentage} ensures that there is no bias against tiny anatomical regions with small number of voxels by normalizing the number of voxels selected from each region by its total number of voxels. Figure \ref{fig:Region_voxel_contribution_Precentage} clearly shows that in the left prefrontal and inferior occipital regions a significant percentage of voxels are active during complex problem solving. Both figures also show high standard deviations across subjects, which indicates high inter-subject variability.

\subsection{Interpolation} 

\begin{figure}[t]
\centering
\includegraphics[width=0.5\textwidth]{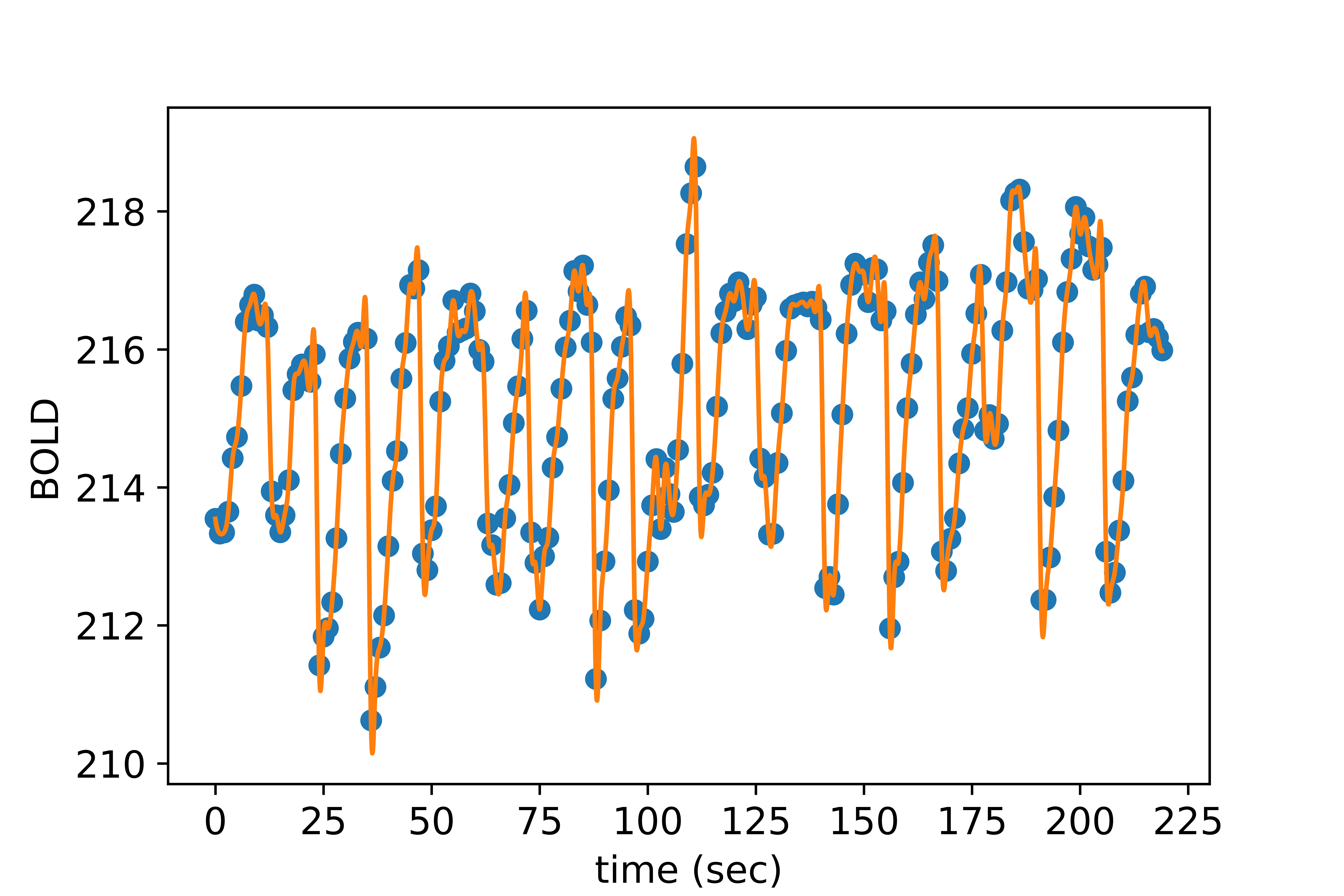}
\caption{Interpolated BOLD response.}
\label{fig:Interpolated_BOLD}
\end{figure}

After selecting the most discriminative voxels and averaging their BOLD responses with respect to their corresponding brain anatomical regions, we employed temporal interpolation to increase the temporal resolution of the TOL dataset. 
As a result, the total number of obtained brain volumes is equal to $ n + z * (n-1)$ where $n$ is the number of measured brain volumes of a given puzzle and $z$ is the number of estimated brain volumes plugged between each pair of measured brain volumes. The optimal value of $z$ is equal to $8$ which is determined empirically using cross-validation. Figure \ref{fig:Interpolated_BOLD} shows the interpolated BOLD response of a randomly selected anatomical region from the given subjects, where the blue dots represent the measured BOLD response of the region and the orange dashes are the interpolated values. It is clear from Figure \ref{fig:Interpolated_BOLD} that the interpolated points using cubic spline function do not introduce sharp edges nor do they smooth out the spikes between measured brain volumes. 

\begin{figure*}[t]
\centering
\includegraphics[width=0.80\textwidth]{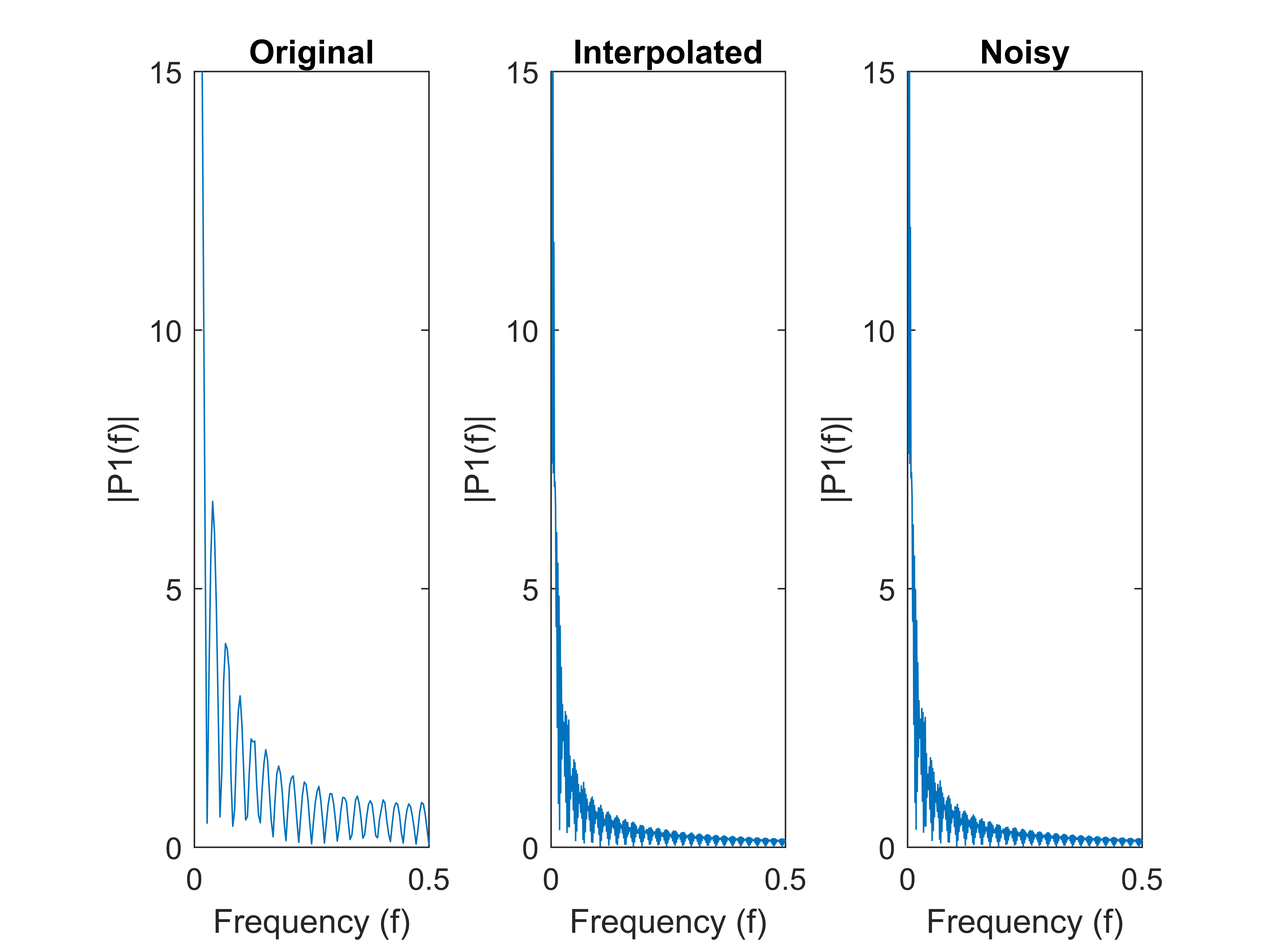}
\caption{Single-Sided amplitude spectrum.}
\label{fig:Amplitude_Spectrum}
\end{figure*}

Furthermore, Figure \ref{fig:Amplitude_Spectrum} shows the single-sided amplitude spectrum of a randomly selected anatomical region from a given subject before interpolation, after interpolation and finally after adding Gaussian noise. The figure clearly demonstrates that both interpolation and injecting Gaussian noise preserve the smooth peaks of the signal in the frequency domain. 

\subsection{Gaussian Noise}

In order to control the signal-to-noise ration (SNR), we used cross-validation to choose the optimal pair of values for $\alpha_{noise}$ and $\beta_{noise} $, the ratios of mean and standard deviation of the added noise respectively. 
As a result, the optimal values obtained are $\alpha_{noise} = 0.025$ and $\beta_{noise} =0.075$ from the following set of values 
$\alpha_{noise}, \beta_{noise} \in [0.025, 0.05, 0.075, 0.1] $. \\

\subsection{Brain Decoding}

We use brain decoding in order to quantify the effect of our proposed preprocessing steps on the TOL dataset. We aim to distinguish the two phases of complex problem solving namely: planning and execution. 
At first, we used ANOVA to select the 10,000 voxels with the highest $f$-scores then we averaged the selected voxels into their corresponding anatomical regions defined by AAL \cite{AAL}. Following that, we employed temporal interpolation to increase the temporal resolution of each puzzle by estimating $z=8$ brain volumes between each pair of measured brain volumes. Finally, we added Gaussian noise in order to regularize the BOLD responses of each region to improve the generalization performance of the classifiers.  
We used $k$-fold Cross validation for each subject in all of the experiments introduced in this section, with $k=8$. After we obtained the results, we averaged them across the different fold, then we calculated the average and standard deviation across all subjects. 
We used both supervised and unsupervised brain decoding methods, a linear support-vector machine (SVM) \cite{fan2008liblinear} was used for supervised brain decoding while $k$-means clustering was used for unsupervised brain decoding. 

Table \ref{table:preprocessing_model} shows the effect of our preprocessing pipeline on the brain decoding of complex problem solving subtasks. The first row shows the performances of brain decoding on the raw dataset without any preprocessing, simply averaging all of the voxels into their corresponding anatomical regions. While the second row shows the results of applying voxel selection then averaging the selected voxels into their anatomical regions. The third row shows the results of brain decoding after applying temporal interpolation, while the forth row shows the results after injecting the data with Gaussian noise.

\begin{table}[h]
\centering
\begin{small}

\begin{tabular}{|l|l|l|}
\cline{1-3}
  \textbf{Preprocessing} & \textbf{SVM} & \textbf{$k$-Means} \\ \cline{1-3}
  Raw data & 0.60 $\pm$ 0.11 & 0.63 $\pm$ 0.09	 \\
  Voxel Selection & 0.74 $\pm$ 0.12 & \textbf{0.85} $\pm$ 0.06	 \\
  Interpolation & 0.81 $\pm$ 0.08 & 0.84 $\pm$ 0.06	 \\
  Noise addition & \textbf{0.82} $\pm$ 0.08 & \textbf{0.85} $\pm$ 0.06 \\ \cline{1-3}

\end{tabular}
\end{small}
\caption{Decoding performances of preprocessing pipeline after each step.}
\label{table:preprocessing_model}
\end{table}

From the results of the preprocessing experiments, it is observed that voxel selection improves the brain decoding performance for both supervised and unsupervised methods from \%60 to \%74 and from \%63 to \%85 respectively. 
This can be attributed to voxel selection retaining only the most discriminative voxels and trashing the remaining less informative ones. In addition, voxel selection manages to sparsify the representation of the data since some brain regions contribute no voxels at all thus have a flat BOLD response.

The table also shows that temporal interpolation further improves the supervised brain decoding performance from \%74 to \%81, this significant increase is due to increasing the number of brain volumes thus increasing the number of training samples for the classifier. However, temporal interpolation slightly reduces the performance of unsupervised methods from \%85 to \%84 which can be partially attributed to the estimated brain volumes during the transitions between the two phases of problem solving, planning and execution which reduces the separation between the two natural subgroups. This is due to the method used to label the estimated brain volumes, where each estimated brain volume is given the labels of its closest neighboring measured brain volume.

Finally, the addition of Gaussian noise slightly boosts the performance of both supervised and unsupervised methods from \%81 to \%82 and from \%84 to \%85 respectively. The table also shows high standard deviation across subjects, which is consistent with voxel selection plots, revealing high inter-subject variability.

\subsection{Building Brain Networks}






In this section, we compare our model for building dynamic functional brain networks with some of the popular methods proposed in the literature in terms of their brain decoding power. Brain decoding can verify whether the constructed brain networks are good representatives of the underlying cognitive subtasks or not. 

For this purpose, we built brain networks as explained in the previous sections after having successfully applied the preprocessing pipeline. The optimal values for learning rate $\alpha_{learning}$ and number of epochs were chosen empirically using cross-validation obtaining the following values respectively $1*10^{-8}$ and 10.
As for $p$, the number of neighbors used to represent each anatomical region, we chose $p$ equal to the total number of regions which is 90, in this way, a fully-connected brain network is obtained at each time window. However, the total number of nodes is less than 90 given that some regions have flat BOLD responses therefore they were pruned along with all their edges from the brain network.

We also constructed brain networks using Pearson correlation and ridge regression as proposed in \cite{richiardi2011decoding,Onal16} and \cite{onal2017new,onal2015modeling} respectively in order to compare the performance of our methods with other works in the literature.
In the case of Pearson correlation, the functional brain networks were constructed using Pearson correlation scores between each pair of brain regions \cite{richiardi2011decoding,Onal16}. As for the case of ridge regression, the mesh arc-weight descriptors were estimated using ridge regression in order to represent each region as a linear combination of its neighbors \cite{onal2017new,onal2015modeling}.

Table \ref{table:proposed_model} shows the brain decoding results of the aforementioned brain network construction methods compared against the results of multi-voxel pattern analysis (MVPA).
The first row shows the brain decoding results of MVPA, while the second and third rows show the results of Pearson correlation and ridge regression methods respectively. The last row shows the brain decoding results of our proposed neural network model. The table clearly shows that both Pearson correlation and ridge regression fail to construct valid brain networks that are good representatives of the underlying cognitive tasks. However, our model managed to get brain decoding results similar or slightly better than those obtained from MVPA both in the cases of supervised and unsupervised methods. This can be attributed to the challenging nature of the TOL dataset, Pearson correlation does not manage to capture the interdependencies between the anatomical regions over short time windows. While ridge regression fails to correctly estimate the mesh arc-weights as it estimates the arc-weights for each region independently of the other ones. Our proposed model, with a relatively small number of epochs manages to obtain mesh arc-weight values that capture the activation patterns of anatomical regions and their relationships.    

\begin{table}[h]
\centering
\begin{small}

\begin{tabular}{ |l |c |c |}
\cline{1-3}
  \textbf{Algorithm} & \textbf{SVM} & \textbf{$k$-Means} \\ \cline{1-3}
  MVPA & \textbf{0.82} $\pm$ 0.08 & 0.85 $\pm$ 0.06	 \\
  Pearson  & 0.58 $\pm$ 0.05 & 0.57 $\pm$ 0.04	 \\
  Ridge Regression & 0.56 $\pm$ 0.05 & 0.55 $\pm$ 0.02	 \\
  Neural Networks & \textbf{0.82} $\pm$ 0.10 & \textbf{0.87} $\pm$ 0.06 \\ \cline{1-3}

\end{tabular}
\end{small}

\caption{Results of proposed model.}
\label{table:proposed_model}
\end{table}

\section{Brain Network Properties}

In this section, we aim to analyze the network properties of the constructed functional brain networks. We investigate the network properties for each anatomical brain region during both planning and execution subtasks in order to understand which regions are most active and which regions work together during each one of the two subtasks of complex problem solving. 

Given that the constructed brain functional networks are both weighted, directed, fully-connected and contain both negative and positive weights, we preprocessed the networks before measuring their network properties. Firstly, we got rid of all the negative weights by shifting all the mesh arc-weights values by a positive quantity equal to the absolute value of the largest negative arc-weight. Then, we normalized the mesh arc-weights to ensure that all of weights are within the range of $[0,1]$. Finally, we measured the network properties on the pruned brain graph, where the brain regions (nodes) contributing no voxels (have a flat BOLD response) and all of their corresponding arc-weights (edges) were deleted from the brain graph. Thus, the networks contained less than 90 regions with their corresponding edges. We used brain connectivity toolbox to calculate the investigated network properties \cite{rubinov2010complex}. 

In order to measure for centrality, the number of neighbors for each anatomical region (P) was chosen to be equal to 89, which is equal to the total number of neighbors for any given node as the total number of brain anatomical regions defined by the AAL atlas \cite{AAL} after deleting the regions residing in the cerebellum equals 90. In addition, since we pruned the nodes that correspond to regions from which no voxels were selected, our constructed brain networks were weighted directed fully-connected networks. Therefore, the in-degree, out-degree and total degree of all nodes in the graph were equal to the total number of anatomical regions retained after voxel selection. 

Therefore, we used node strength and node betweenness centrality to identify nodes with high centrality which are potential hubs in the brain networks controlling the flow of information in the network. 
In our proposed model, the node in-strength of node $i$ is the sum of the mesh arc-weight values which is estimated using our proposed neural network method in order to minimize the reconstruction error of the BOLD response of anatomical region $i$ using its neighbors. Thus, node in-strength is not used as part of our network properties analyses, we rather used node out-strength to measure the centrality of all anatomical regions. 

As for measures of segregation, quantifying the existence of subgroups within brain networks is based on densely interconnected nodes. These subgroups are commonly referred to as clusters or modules. The existence of such clusters in functional brain networks is a sign of interdependence among the nodes forming the cluster. Therefore, clustering coefficient, transitivity and local efficiency were measured in order to identify potential clusters with dense interconnections in the brain networks.

\subsection{Planning \& Execution Brain Networks}

In this section, we discuss the network properties of the planning and execution networks. For each aforementioned network metric, we ranked the brain regions in descending order according to their score on that network measure for all subjects across all sessions. Then, we retained the 10 anatomical regions with the highest scores. Following that, we measured the frequency of occurrence of each brain region among the top 10 regions across all sessions in order to identify the shared regions and patterns across all subjects for both planning and execution subtasks. The results of the analysis are shown in tables: table \ref{table:Absolute_Activation_Planning} shows the brain regions that have high scores for the reported network properties during planning subtask, and table \ref{table:Absolute_Activation_Execution} shows the brain regions that have high scores during execution subtask.

\begin{table*}[h]
\centering
\begin{small}
\begin{tabular}{ |c| c| c| c| c| }
\cline{1-5}
 \textbf{transitivity} 		& \textbf{local efficiency}  & \textbf{clustering coefficient}  &  \textbf{betweenness} 		  & \textbf{out-strength} \\ \cline{1-5}
 \color{blue} Angular 			& Calcarine 		& Calcarine  			  &  \color{red}Cueneus R 			  & Cueneus R	 \\
 \color{blue}Calcarine 			& \color{purple}Cuneus  			& \color{magenta}Cuneus  				  &  \color{red}Frontal Sup L		  & Frontal Sup L  \\
 \color{blue}Cingulum Ant 		& Frontal Mid R 	& \color{magenta}Frontal Mid R 		  &	 \color{red}Fusiform R			  & Fusiform R \\
 \color{blue}Cingulum Mid 		& Frontal Sup		& \color{magenta}Frontal Sup   		  &  \color{red}Paracentral Lobule L & \color{orange}Paracentral Lobule L\\
\color{blue}Cuneus  			& \color{purple}Fusiform 			& \color{magenta}Fusiform 				  &  Parietal Sup R       & \color{orange}Supp Motor Area R \\
\color{blue} Frontal Inf Oper L & \color{purple}Occipital Inf R   & \color{magenta}Occipital Inf R  		  &	 \color{red}Precuneus L		  & \color{orange}Temporal Inf R\\
 					& Precentral 		& \color{magenta}Parietal Sup R 		  &  \color{red}Supp Motor Area R    & \color{orange}Temporal Mid R\\
   					& \color{purple}Supp Motor Area R	& \color{magenta}Precentral 			  &  \color{red}Temporal Inf R 	  &    \\
   					& \color{purple}Temporal Inf R	& \color{magenta}Supp Motor Area R 	  &  \color{red}Temporal Mid R  	 &  \\
   					&					& \color{magenta}Temporal Inf R 		  &   					 &  \\ \cline{1-5}
\end{tabular}
\end{small}
\caption{\textbf{Planning}: Anatomical regions with the highest network measures across subjects, regions are painted if they overlap with execution.}
\label{table:Absolute_Activation_Planning}
\end{table*}

There are a number of processes taking place during planning and execution. Plan generation involves a series of recursive events including: 1) problem encoding; 2) decision-making in order to decide which ball to move and where to move it; 3) mental imagery to imagine the ball moving; and 4) working memory to maintain the intermediate steps as well as the move number. During plan execution there is 1) retrieval of the steps from memory; 2) confirming the correct steps are being performed; and 3) the motor execution of those steps. As the results demonstrate the networks for planning and execution are overlapping. These results are similar to the activation results reported in \cite{Newman2009} in that the regions that were found to be active during the task are also regions that are most prominently found with the highest network measures.  These regions include the right and left middle frontal gyrus, anterior cingulate cortex, precentral cortex, and superior parietal cortex.  

Previous work has suggested that the regions found in the current study to show high network measures are directly related to the sub-tasks associated with TOL performance. For example, both the left and right prefrontal cortex have been found to be involved in the TOL task with the two regions performing distinguishable functions. The right prefrontal cortex is involved in constructing the plan for solving the TOL problem while the left prefrontal cortex is involved in supervising the execution of that plan \cite{Newman2009,Newman2003}. The anterior cingulate has been linked to error detection and is particularly involved in the TOL when the number of moves is higher or the problem difficulty is manipulated. The right superior parietal cortex and precentral cortex have been linked to visuo-spatial attention necessary for planning \cite{Newman2003} and the left parietal cortex has been linked to visuo-spatial working memory processing \cite{Newman2003}. The overlap between the regions with the highest network measures and those that have been linked to the task is an important feature and is not due to the voxel selection process. Many regions that passed threshold were not in the top ranked list of network measures. For example, the basal ganglia including the caudate has been found in previous studies to be involved in TOL performance \cite{Newman2009,beauchamp2003dynamic,dagher1999mapping,rowe2001imaging,van2003frontostriatal}; however, the region appears to not be an important network hub. 

\begin{table*}[h]
\centering
\begin{small}
\begin{tabular}{ |c| c| c| c| c| }
\cline{1-5}
 \textbf{transitivity} 	  & \textbf{local efficiency}  & \textbf{clustering coefficient}  &  \textbf{betweenness} 			& \textbf{out-strength} \\ \cline{1-5}
 \color{blue}Angular 		  & Calcarine L 	  & Calcarine L  			&  \color{red}Cueneus R     		& Frontal Sup	 \\
 \color{blue}Calcarine 		  & \color{purple}Cuneus  		  & \color{magenta}Cuneus   				&  \color{red}Frontal Sup L    	& Fusiform \\
 \color{blue}Cingulum Ant 	  & Frontal Sup  L    & \color{magenta}Frontal Mid R 			&  \color{red}Fusiform R      		& \color{orange}Paracentral Lobule L \\
 \color{blue}Cingulum Mid 	  & \color{purple}Fusiform 		  & \color{magenta}Frontal Sup 	  		&  \color{red}Paracentral Lobule L & \color{orange}Supp Motor Area R\\
 \color{blue}Cuneus 		  & \color{purple}Occipital Inf R   & \color{magenta}Fusiform 				&  \color{red}Precuneus L 			& \color{orange}Temporal Inf R \\
 \color{blue}Frontal Inf Oper & \color{purple}Supp Motor Area R & \color{magenta}Occipital Inf R  		&  \color{red}Supp Motor Area R    & \color{orange}Temporal Mid R\\
 				  & \color{purple}Temporal Inf R 	  & \color{magenta}Parietal Sup R 			&  \color{red}Temporal Inf R		&   \\
   			      &					  & \color{magenta}Precentral R  			&  \color{red}Temporal Mid R       &    \\
   			      &					  & \color{magenta}Supp Motor Area R 		&   					&  \\
   			      &					  & \color{magenta}Temporal Inf R			&       				&  \\ \cline{1-5}
\end{tabular}
\end{small}
\caption{\textbf{Execution}: Anatomical regions with the highest network measures across subjects, regions are painted if they overlap with planning.}
\label{table:Absolute_Activation_Execution}
\end{table*}

\begin{figure*}[h]
		\centering
        \begin{subfigure}{\textwidth}
    	\centering
\includegraphics[width=0.75\textwidth]{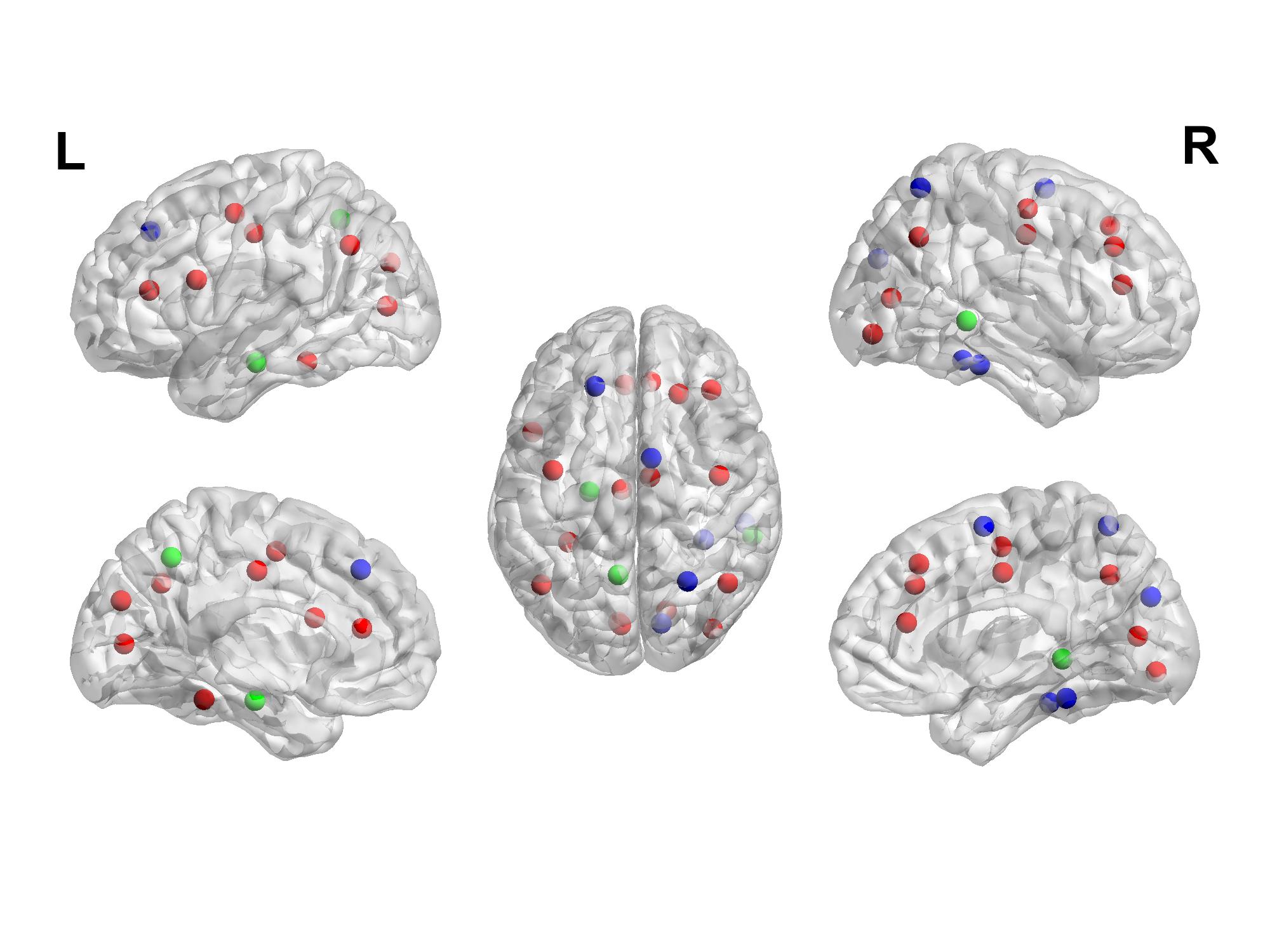}
\caption{Planning Brain Network.}
\label{fig:planning_absolute}
        \end{subfigure}
        \centering
        \begin{subfigure}{\textwidth}
    	\centering
\includegraphics[width=0.75\textwidth]{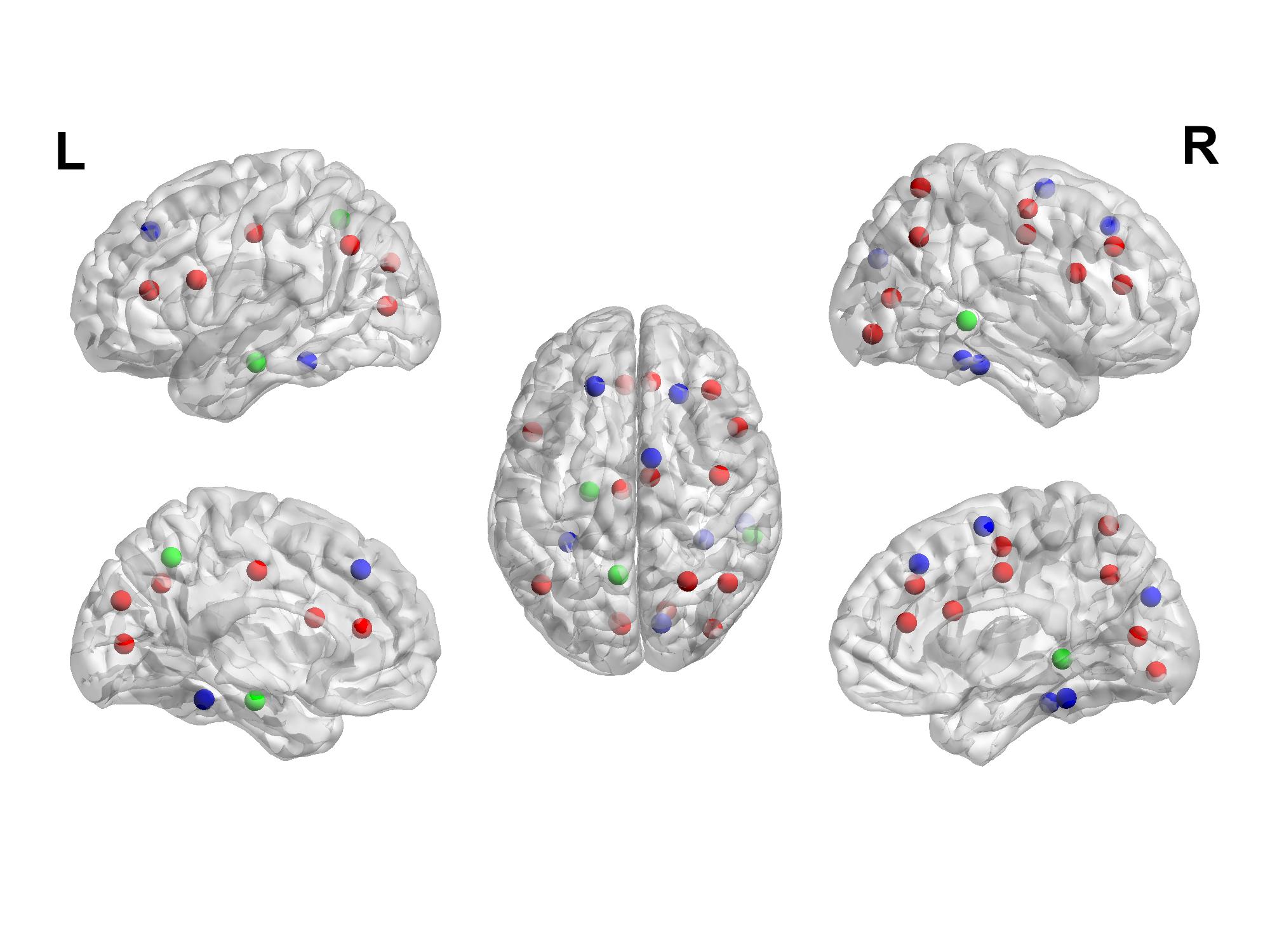}
\caption{Execution Brain Network.}
\label{fig:execution_absolute}
        \end{subfigure}
\caption{Regions with the highest network measures for Planning (Top) and Execution (Bottom).}
\end{figure*}

Figures \ref{fig:planning_absolute} and \ref{fig:execution_absolute} visualize the reported brain regions in tables
\ref{table:Absolute_Activation_Planning} and \ref{table:Absolute_Activation_Execution} respectively using Brain Net Viewer \cite{brainnet}.
In figures \ref{fig:planning_absolute} and \ref{fig:execution_absolute}, the color of the node (brain region) implies the following: red indicates that the region has high transitivity, clustering coefficient or local efficiency. Green indicates that the node has high node centrality measured by node out-strength and node betweenness. As for blue, it shows the nodes that have high node centrality and is part of subgroup of densely interconnected regions.

\subsection{Differences between Planning and Execution Networks}

In this section, we explore the network differences between planning and execution by calculating the difference between the network property scores for planning and execution for each session. To achieve that, we took the difference between the network property scores for brain anatomical regions during planning and the network property scores for brain anatomical regions during execution for each session. Then, we counted the frequency of times a given anatomical region is more active during planning than execution and vice-versa in order to identify consistent patterns of the disagreements between planning brain networks and execution brain networks across all subjects. Results showed, generally, that the network measures were higher for planning than execution. This, too, mirrors the findings from \cite{Newman2009} in which planning resulted in greater activation than execution.

Node out-strength is a measure of how connected the node is to other nodes in the network. Planning showed greater out-strength than  execution in the following regions: occipital regions (calcarine, cuneus), parietal regions (bilateral superior parietal cortex and precunues), the right superior frontal cortex, and inferior occipito-temporal regions (fusiform and lingual gyri). The left angular gyrus and bilateral medial superior frontal cortex showed greater out-strength for execution. As for node betweenness, the following brain regions had higher node betweenness during planning than execution: occipital regions (calcarine, cuneus, right middle, right superior); inferior occipito-temporal (fusiform, lingual); parietal (bilateral superior parietal, left postcentral, precuneus). Bilateral medial superior frontal had higher node betweenness during execution than planning.  

These results suggest that there is greater information flow during planning than execution. This matches our expectations. Planning is more computationally demanding than execution. Again, during planning participants must explore the problem space which requires generating and manipulating a mental representation of the problem. The regions that show greater information flow during planning are all regions involved in that generation and manipulation particularly parietal, occipital and inferior occipito-temporal. On the other hand, execution requires recall of the plan generated and stored and therefore, greater information flow from frontal regions related to memory retrieval is observed.  

Clustering coefficient, local efficency and transitvity are measures of segregation which aim to identify sub-networks. Each of these measures were larger for planning than execution with no regions showing larger measures for execution. The regions that showed higher clustering coefficient in planning included: the cuneus, left middle occipital cortex, and right precuneus. Local efficiency was higher in a similar set of regions (the cuneus, left middle occipital cortex, and right precuneus). The clustering coeffiencent and local  efficiency identified a visual-spatial sub-network that is more strongly connected during planning. Transitivity identified an overlapping but more extensive set of regions that included: bilateral angular gyrus, calcarine sulcus, cuneus, bilateral middle frontal cortex, bilaterial superior frontal cortex, bilateral fusiform and lingual gyri, bilateral occipital cortex, bilatral superior parietal cortex, postcentral and precentral cortex, precuneus, supplementary motor area, right supramarginal gyrus, and right inferior and middle temporal cortex.

\begin{figure*}[h]
		\centering
        \begin{subfigure}{\textwidth}
    	\centering
\includegraphics[width=0.75\textwidth]{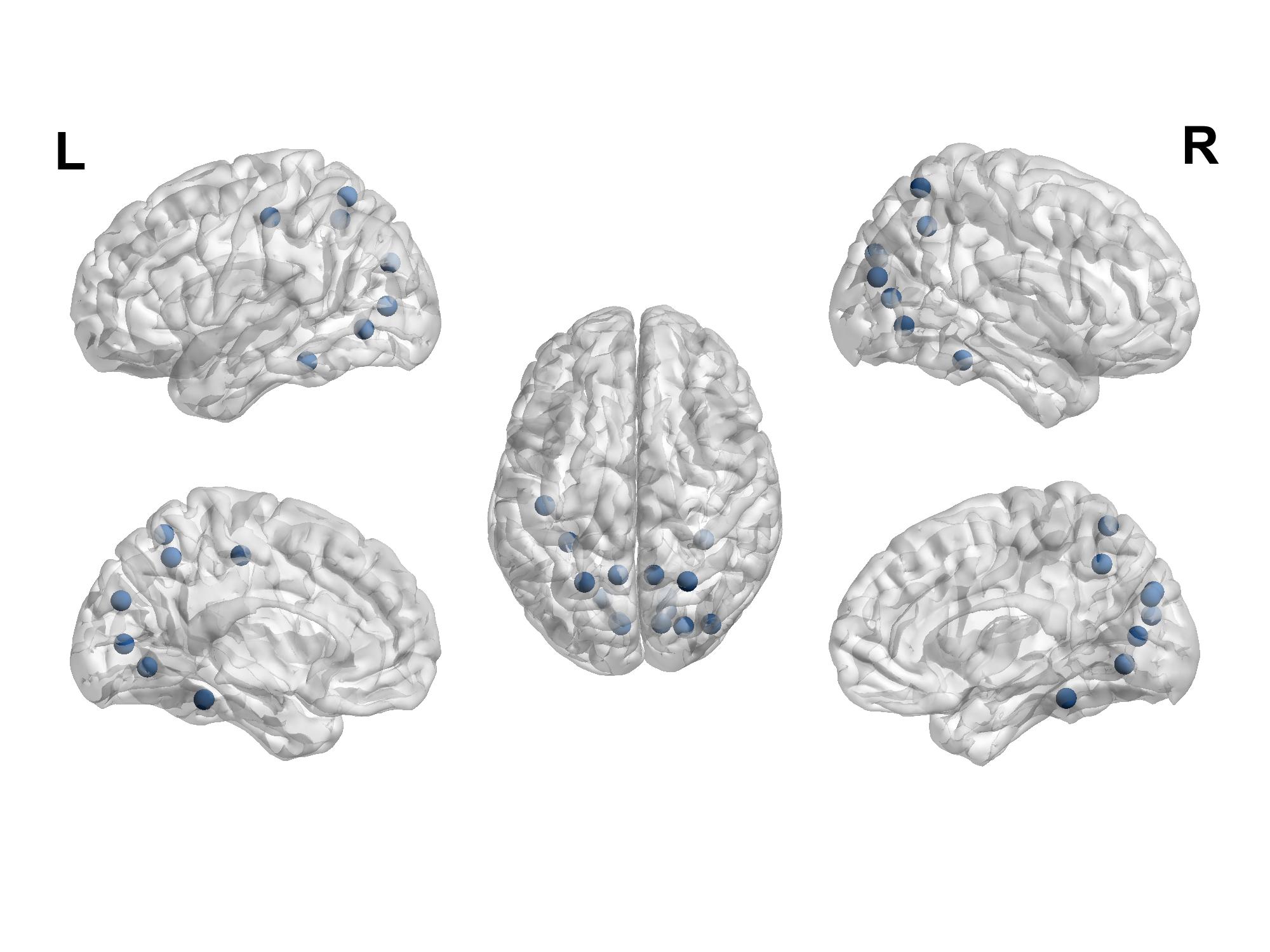}
\caption{Anatomical regions with higher node betweenness during planning.}
\label{fig:planning_more_betweeness}
        \end{subfigure}
        \centering
        \begin{subfigure}{\textwidth}
    	\centering
\includegraphics[width=0.75\textwidth]{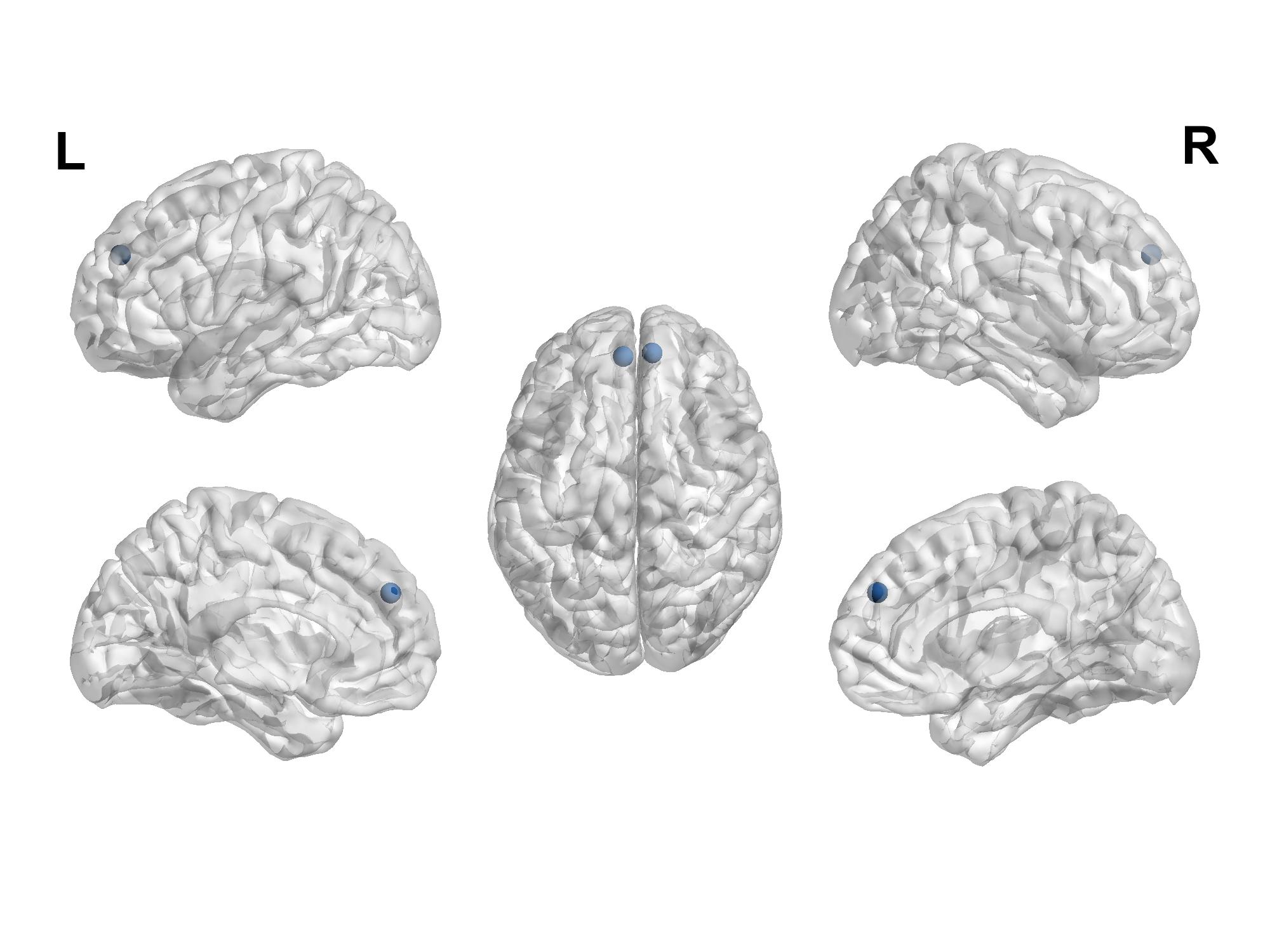}
\caption{Anatomical regions with higher node betweenness during execution.}
\label{fig:execution_more_betweeness}
        \end{subfigure}
\caption{Anatomical regions with higher node betweenness during planning (Top) and during execution (Bottom).}

\end{figure*}

Figures \ref{fig:planning_more_betweeness} , \ref{fig:execution_more_betweeness} visualize the brain regions with higher betweenness during planning and during execution respectively.

\begin{figure*}[h]
		\centering
        \begin{subfigure}{\textwidth}
    	\centering
\includegraphics[width=0.75\textwidth]{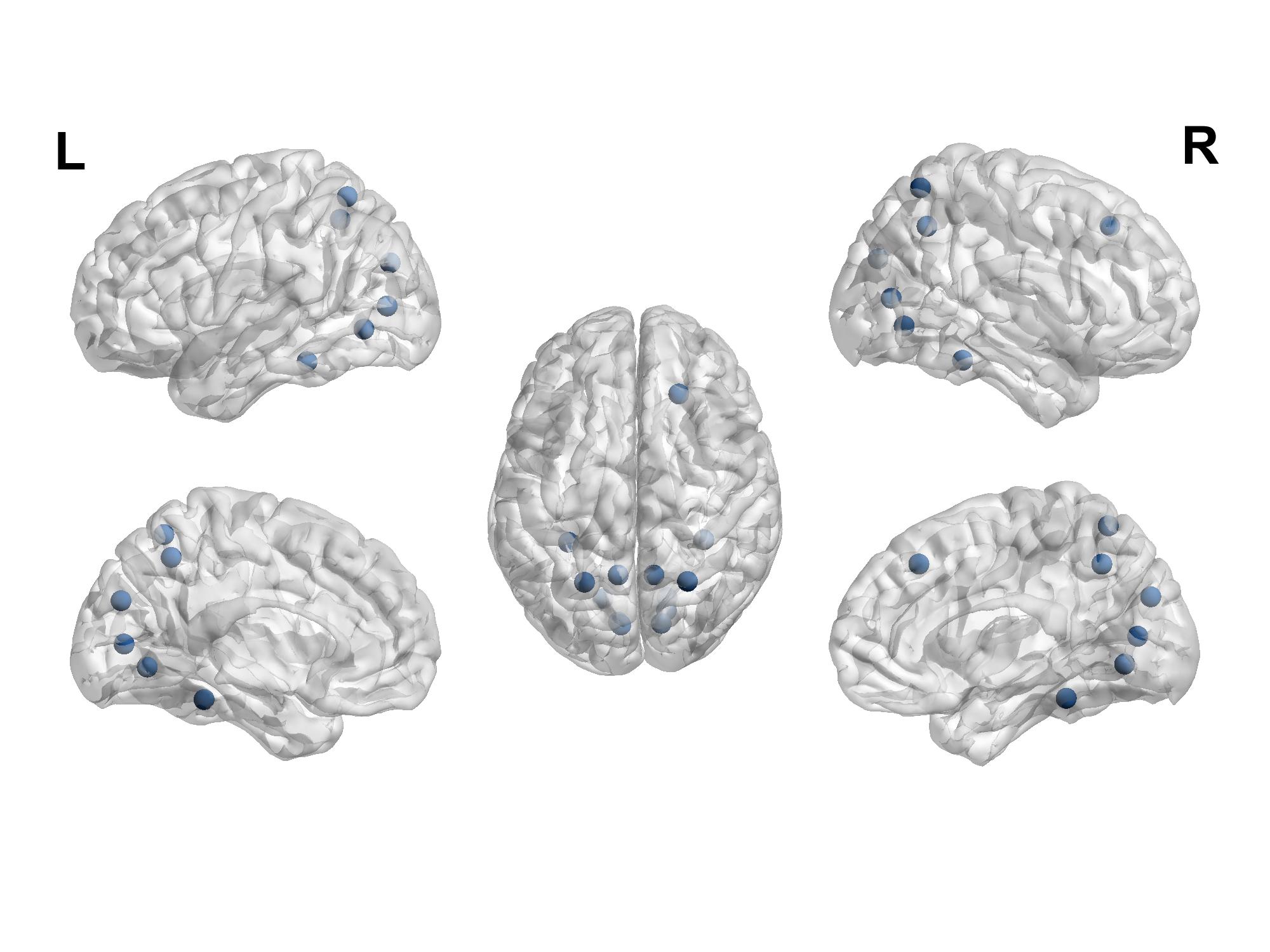}
\caption{Anatomical regions with higher node out-strength during planning.}
\label{fig:planning_more_out_strength}
        \end{subfigure}
        \centering
        \begin{subfigure}{\textwidth}
    	\centering
\includegraphics[width=0.75\textwidth]{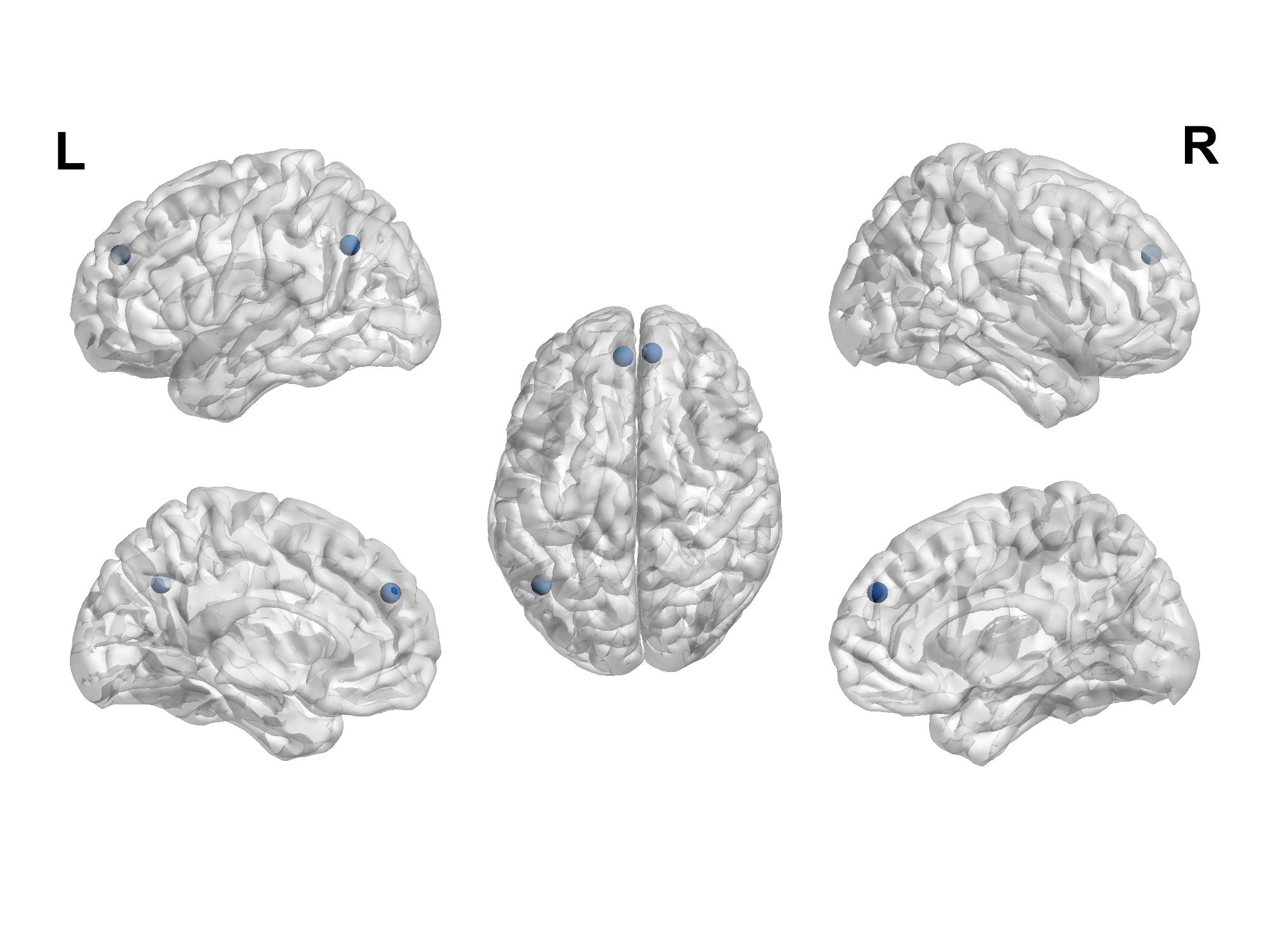}
\caption{Anatomical regions with higher node out-strength during execution.}
\label{fig:execution_more_out_strength}
        \end{subfigure}
\caption{Anatomical regions with higher node out-strength during planning (Top) and during Execution (Bottom).}
\end{figure*}

Figures \ref{fig:planning_more_out_strength} , \ref{fig:execution_more_out_strength} visualize the brain regions with higher node out-strength during planning and during execution, respectively.

\begin{figure*}[h]
		\centering
        \begin{subfigure}{\textwidth}
    	\centering
\includegraphics[width=0.75\textwidth]{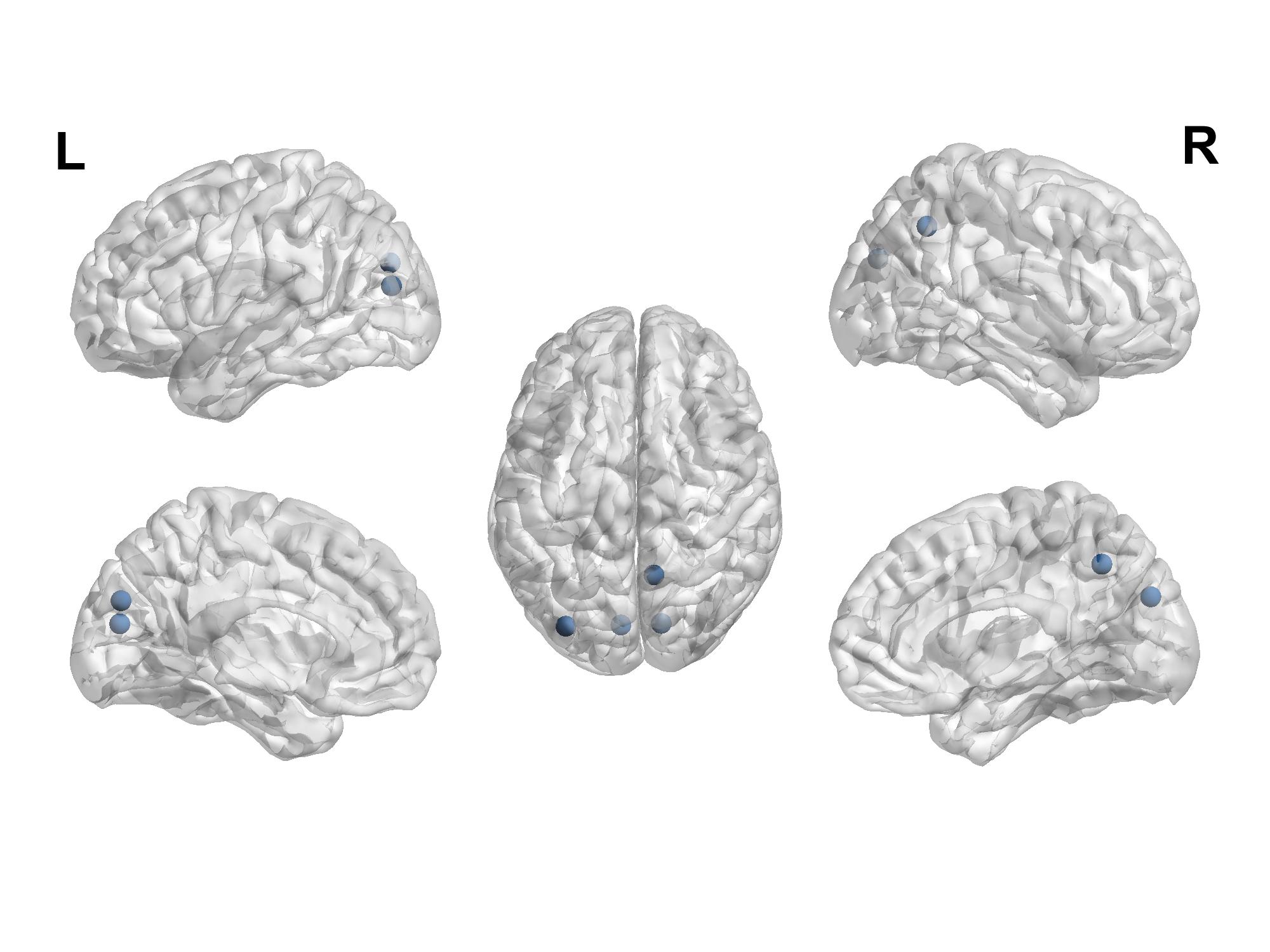}
\caption{Anatomical regions with higher local efficiency and clustering coefficient during planning.}
\label{fig:planning_more_local_eff_and_clustering_coef}
        \end{subfigure}
        \centering
        \begin{subfigure}{\textwidth}
    	\centering

\includegraphics[width=0.75\textwidth]{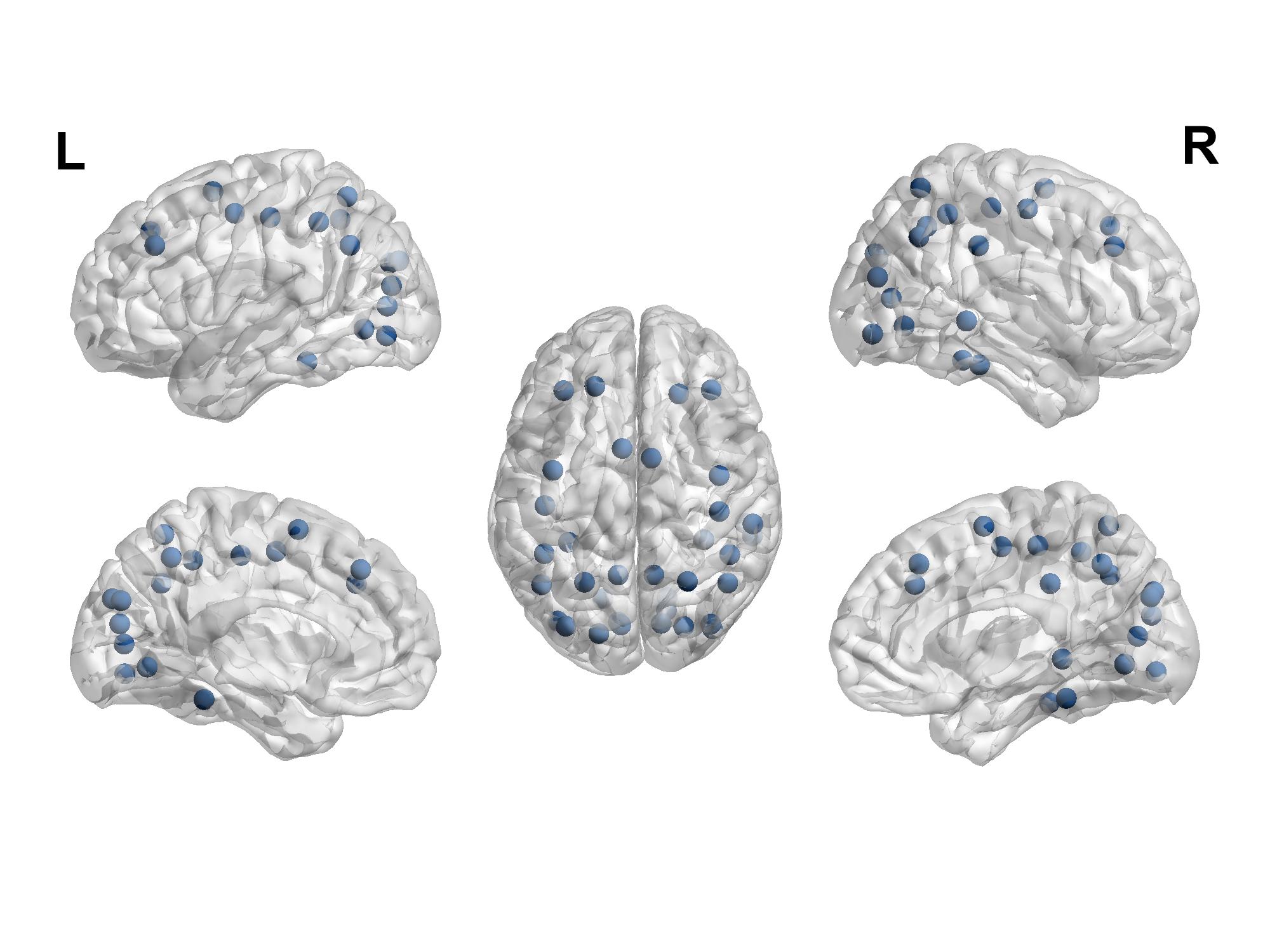}
\caption{Anatomical regions with higher transitivity during planning.}
\label{fig:planning_more_transitivity}
        \end{subfigure}
\caption{Anatomical regions with higher local efficiency and clustering coefficient (Top) and higher transitivity (Bottom) during planning.}
\end{figure*}

Figure \ref{fig:planning_more_local_eff_and_clustering_coef} visualizes the brain regions with higher local efficiency and higher clustering coefficient during planning phase compared to execution phase. While Figure \ref{fig:planning_more_transitivity} visualizes the brain regions with higher transitivity during planning than during execution phase.

\subsection{Global Efficiency}

Since global efficiency is measured over the entire brain network, not for a given node in the network, we measured the global efficiency for all planning and execution networks within all sessions across subjects. Then, global efficiency of planning is compared against that of execution. Results show that the majority of sessions had higher global efficiency scores during planning than execution, 43 out of 72 sessions had higher global efficiency during planning than execution. Furthermore, table \ref{table:Global_Efficiency} shows the number of sessions where global efficiency was higher during planning and during execution across all subjects for all 4 sessions of each subject. The first column shows the number of subjects that had a higher global efficiency score during planning than during execution. The second column shows the number of subjects that had a higher global efficiency score during execution than during planning. 

Although there was no significant difference in global efficiency between planning and execution, from the table, it is clear that the majority of subjects had a higher global efficiency for planning for the first session. Some subjects switched from having higher global efficiency during planning to having higher global efficiency during execution. A potential explanation for this change across sessions is a switch from pre-planning to on-line planning, or planning intermixed with execution. Although there is a dedicated planning phase in the current study, that does not mean that planning is not taking place during execution. In fact, it has been debated as to whether efficient pre-planning is possible in the TOL or whether TOL performance is controlled by on-line planning \cite{phillips1999role,phillips2001research,kafer1997testing,Unterrainer2004}. According to Phillips et.al. \cite{phillips1999role,phillips2001research} pre-planning the entire sequence is not natural, but that people instead plan the beginning sequence of moves and then intersperse planning and execution. If this is the case then it may be expected that some participants will switch to on-line planning. This intermixing of planning and execution is also likely to impact the performance of the machine learning algorithms to detect planning and execution phases.

The relationship between global efficiency and behavioral performance was examined. Global efficiency was found to be positively correlated with the mean number of extra moves (a measure of error) during problem-solving (for execution r=0.73, p=0.0006). Previous studies have shown a relationship between global efficiency and task performance \cite{stanley2015changes}. 

This suggests that the variance in global efficiency is indicative of individual differences in neural processing and further suggests that the changes in global efficiency across sessions are also likely indicative of changes in neural processing related to changing strategy. Further research using a larger sample is necessary to explore this hypothesis.

\begin{table}[h]
\centering
\begin{small}

\begin{tabular}{ |c| c| c| }
\cline{1-3}

  \textbf{Session Number}  & \textbf{Planning} & \textbf{Execution} \\ \cline{1-3}
   1 & 15 & 3	 \\
   2  & 9 & 9	 \\
   3 & 10 & 8	 \\
   4 & 9 & 9 \\ \cline{1-3}

\end{tabular}
\end{small}

\caption{Global Efficiency.}
\label{table:Global_Efficiency}
\end{table}

\section{Conclusion}

In this paper, we proposed a model to construct brain functional networks during a complex problem solving task. Our model successfully identified the two phases of complex problem solving. In addition, the network properties of the constructed brain networks during planning and execution phases were studied in order to identify essential nodes within the brain networks related to problem solving, potential hubs, and densely connected clusters. Furthermore, the differences between planning networks and execution networks were highlighted and discussed.

There are some limitations to the study. Although the primary aim of this study was to demonstrate the feasibility of the methods, the sample size is somewhat small, making the interpretation of the results difficult. Second, a goal of this method is to identify brain states that are interspersed with each other. In the current study planning was expected to occur both prior to execution as well as during execution therefore planning states are interspersed within the execution phase. The temporal sampling rate of the fMRI data may be a limiting factor. Alternatively, the sluggish and blurred underlying hemodynamic response may be the factor preventing the ability to detect brain states. We plan to explore this factor in future work.

\section{Acknowledgment}
The work is supported by TUBITAK (Scientific and Technological Research Council of Turkey) under the grant No: 116E091 as well as the Indiana METACyt Initiative of Indiana University, funded in part through a major grant from the Lilly Endowment, Inc.

\bibliographystyle{IEEEbib}
\bibliography{strings}

\end{document}